\pdfoutput=1

\documentclass[12pt,a4paper]{article}

\usepackage{ifthen} 
\newboolean{pdflatex}
\setboolean{pdflatex}{false} 

\newboolean{articletitles}
\setboolean{articletitles}{true} 

\newboolean{uprightparticles}
\setboolean{uprightparticles}{false} 

\newboolean{inbibliography}
\setboolean{inbibliography}{false} 

\def\paperauthors{LHCb collaboration} 
\def\paperasciititle{Amplitude analysis of the decay B0bar to K0S pi pi and first observation of CP asymmetry in the transition B0bar to K*(892)- pi+} 
\def\papertitle{Amplitude analysis  of the decay $\Bdb \to K_{\mathrm{\scriptscriptstyle S}}^0 \pi^+ \pi^-$ and first observation of the \CP asymmetry in $\Bdb \to K^{*}(892)^- \pi^+$} 
\def\paperkeywords{{High Energy Physics}, {LHCb}} 
\def\papercopyright{CERN on behalf of the LHCb collaboration}
\def\paperlicence{CC-BY-4.0}
\def\paperlicenceurl{https://creativecommons.org/licenses/by/4.0/}

\usepackage[top=1in, bottom=1.25in, left=1in, right=1in]{geometry}

\usepackage{microtype}
\usepackage{lineno}  
\usepackage{xspace} 
\usepackage{caption} 

\usepackage{graphicx}  
\usepackage{color}
\usepackage{colortbl}
\graphicspath{{./figs/}} 

\usepackage{amsmath} 
\usepackage{amssymb}
\usepackage{amsfonts}
\usepackage{upgreek} 

\newcommand*\patchAmsMathEnvironmentForLineno[1]{%
\expandafter\let\csname old#1\expandafter\endcsname\csname #1\endcsname
\expandafter\let\csname oldend#1\expandafter\endcsname\csname
end#1\endcsname
 \renewenvironment{#1}%
   {\linenomath\csname old#1\endcsname}%
   {\csname oldend#1\endcsname\endlinenomath}%
}
\newcommand*\patchBothAmsMathEnvironmentsForLineno[1]{%
  \patchAmsMathEnvironmentForLineno{#1}%
  \patchAmsMathEnvironmentForLineno{#1*}%
}
\AtBeginDocument{%
\patchBothAmsMathEnvironmentsForLineno{equation}%
\patchBothAmsMathEnvironmentsForLineno{align}%
\patchBothAmsMathEnvironmentsForLineno{flalign}%
\patchBothAmsMathEnvironmentsForLineno{alignat}%
\patchBothAmsMathEnvironmentsForLineno{gather}%
\patchBothAmsMathEnvironmentsForLineno{multline}%
\patchBothAmsMathEnvironmentsForLineno{eqnarray}%
}

\usepackage{hyperxmp}

\usepackage[pdftex,
            pdfauthor={\paperauthors},
            pdftitle={\paperasciititle},
            pdfkeywords={\paperkeywords},
            pdfcopyright={Copyright (C) \papercopyright},
            pdflicenseurl={\paperlicenceurl}]{hyperref}

\usepackage[all]{hypcap} 

\usepackage{xspace} 
\usepackage{upgreek}

\def\lhcb {\mbox{LHCb}\xspace}

\def\babar  {\mbox{BaBar}\xspace}
\def\belle  {\mbox{Belle}\xspace}

\def\MagUp {\mbox{\em Mag\kern -0.05em Up}\xspace}

\ifthenelse{\boolean{uprightparticles}}%
{

 \def\Peta        {\ensuremath{\upeta}\xspace}

 \def\Ppi         {\ensuremath{\uppi}\xspace}

 \def\Ppsi        {\ensuremath{\uppsi}\xspace}

 \def\PDelta      {\ensuremath{\Delta}\xspace}                 
 \def\PXi      {\ensuremath{\Xi}\xspace}                 
 \def\PLambda      {\ensuremath{\Lambda}\xspace}                 
 \def\PSigma      {\ensuremath{\Sigma}\xspace}                 
 \def\POmega      {\ensuremath{\Omega}\xspace}                 
 \def\PUpsilon      {\ensuremath{\Upsilon}\xspace}                 
 
 \def\PB      {\ensuremath{\mathrm{B}}\xspace}                 
                  
 \def\PD      {\ensuremath{\mathrm{D}}\xspace}

 \def\PJ      {\ensuremath{\mathrm{J}}\xspace}                 
 \def\PK      {\ensuremath{\mathrm{K}}\xspace}

 \def\Pb      {\ensuremath{\mathrm{b}}\xspace}                 
 \def\Pc      {\ensuremath{\mathrm{c}}\xspace}

 \def\Pi      {\ensuremath{\mathrm{i}}\xspace}

 \def\Pq      {\ensuremath{\mathrm{q}}\xspace}                 
                  
 \def\Ps      {\ensuremath{\mathrm{s}}\xspace}

}
{

 \def\Peta        {\ensuremath{\eta}\xspace}

 \def\Ppi         {\ensuremath{\pi}\xspace}

 \def\Ppsi        {\ensuremath{\psi}\xspace}                 
                  
 \mathchardef\PDelta="7101
 \mathchardef\PXi="7104
 \mathchardef\PLambda="7103
 \mathchardef\PSigma="7106
 \mathchardef\POmega="710A
 \mathchardef\PUpsilon="7107
                  
 \def\PB      {\ensuremath{B}\xspace}                 
                  
 \def\PD      {\ensuremath{D}\xspace}

 \def\PJ      {\ensuremath{J}\xspace}                 
 \def\PK      {\ensuremath{K}\xspace}

 \def\Pb      {\ensuremath{b}\xspace}                 
 \def\Pc      {\ensuremath{c}\xspace}

 \def\Pi      {\ensuremath{i}\xspace}

 \def\Pq      {\ensuremath{q}\xspace}                 
                  
 \def\Ps      {\ensuremath{s}\xspace}

}

\makeatletter
\ifcase \@ptsize \relax
  \newcommand{\miniscule}{\@setfontsize\miniscule{4}{5}}
\or
  \newcommand{\miniscule}{\@setfontsize\miniscule{5}{6}}
\or
  \newcommand{\miniscule}{\@setfontsize\miniscule{5}{6}}
\fi
\makeatother

\DeclareRobustCommand{\optbar}[1]{\shortstack{{\miniscule (\rule[.5ex]{1.25em}{.18mm})}
  \\ [-.7ex] $#1$}}

\def\quark     {{\ensuremath{\Pq}}\xspace}
\def\quarkbar  {{\ensuremath{\overline \quark}}\xspace}

\def\squark    {{\ensuremath{\Ps}}\xspace}

\def\cquark    {{\ensuremath{\Pc}}\xspace}
\def\cquarkbar {{\ensuremath{\overline \cquark}}\xspace}

\def\bquark    {{\ensuremath{\Pb}}\xspace}

\def\pion   {{\ensuremath{\Ppi}}\xspace}

\def\pip    {{\ensuremath{\pion^+}}\xspace}
\def\pim    {{\ensuremath{\pion^-}}\xspace}

\def\pimp   {{\ensuremath{\pion^\mp}}\xspace}

\def\kaon    {{\ensuremath{\PK}}\xspace}
  \def\Kbar    {{\kern 0.2em\overline{\kern -0.2em \PK}{}}\xspace}

\def\KorKbar    {\kern 0.18em\optbar{\kern -0.18em K}{}\xspace}
\def\Kz      {{\ensuremath{\kaon^0}}\xspace}
\def\Kzb     {{\ensuremath{\Kbar{}^0}}\xspace}
\def\Kp      {{\ensuremath{\kaon^+}}\xspace}
\def\Km      {{\ensuremath{\kaon^-}}\xspace}
\def\Kpm     {{\ensuremath{\kaon^\pm}}\xspace}

\def\KS      {{\ensuremath{\kaon^0_{\mathrm{ \scriptscriptstyle S}}}}\xspace}

\newcommand{\etapr}{\ensuremath{\Peta^{\prime}}\xspace}

  \def\Dbar    {{\kern 0.2em\overline{\kern -0.2em \PD}{}}\xspace}
\def\D       {{\ensuremath{\PD}}\xspace}

\def\DorDbar    {\kern 0.18em\optbar{\kern -0.18em D}{}\xspace}
\def\Dz      {{\ensuremath{\D^0}}\xspace}

\def\Dstarp  {{\ensuremath{\D^{*+}}}\xspace}

\def\B       {{\ensuremath{\PB}}\xspace}
\def\Bbar    {{\ensuremath{\kern 0.18em\overline{\kern -0.18em \PB}{}}}\xspace}

\def\BorBbar    {\kern 0.18em\optbar{\kern -0.18em B}{}\xspace}

\def\Bd      {{\ensuremath{\B^0}}\xspace}
\def\Bs      {{\ensuremath{\B^0_\squark}}\xspace}
\def\Bsb     {{\ensuremath{\Bbar{}^0_\squark}}\xspace}
\def\Bdb     {{\ensuremath{\Bbar{}^0}}\xspace}

\def\jpsi     {{\ensuremath{{\PJ\mskip -3mu/\mskip -2mu\Ppsi\mskip 2mu}}}\xspace}

  \def\Y#1S{\ensuremath{\PUpsilon{(#1S)}}\xspace}

\def\Lbar        {{\ensuremath{\kern 0.1em\overline{\kern -0.1em\PLambda}}}\xspace}
\def\LorLbar    {\kern 0.18em\optbar{\kern -0.18em \PLambda}{}\xspace}

\newcommand{\decay}[2]{\ensuremath{#1\!\to #2}\xspace}         

\def\to                 {\ensuremath{\rightarrow}\xspace}

\def\CP                {{\ensuremath{C\!P}}\xspace}

\def\AT#1     {\ensuremath{A_{\mathrm{T}}^{#1}}\xspace}           

\def\C#1      {\ensuremath{\mathcal{C}_{#1}}\xspace}                       
\def\Cp#1     {\ensuremath{\mathcal{C}_{#1}^{'}}\xspace}                    
\def\Ceff#1   {\ensuremath{\mathcal{C}_{#1}^{\mathrm{(eff)}}}\xspace}        
\def\Cpeff#1  {\ensuremath{\mathcal{C}_{#1}^{'\mathrm{(eff)}}}\xspace}       
\def\Ope#1    {\ensuremath{\mathcal{O}_{#1}}\xspace}                       
\def\Opep#1   {\ensuremath{\mathcal{O}_{#1}^{'}}\xspace}                    

\newcommand{\tev}{\ifthenelse{\boolean{inbibliography}}{\ensuremath{~T\kern -0.05em eV}}{\ensuremath{\mathrm{\,Te\kern -0.1em V}}}\xspace}
\newcommand{\gev}{\ensuremath{\mathrm{\,Ge\kern -0.1em V}}\xspace}
\newcommand{\mev}{\ensuremath{\mathrm{\,Me\kern -0.1em V}}\xspace}
\newcommand{\kev}{\ensuremath{\mathrm{\,ke\kern -0.1em V}}\xspace}
\newcommand{\ev}{\ensuremath{\mathrm{\,e\kern -0.1em V}}\xspace}
\newcommand{\gevc}{\ensuremath{{\mathrm{\,Ge\kern -0.1em V\!/}c}}\xspace}
\newcommand{\mevc}{\ensuremath{{\mathrm{\,Me\kern -0.1em V\!/}c}}\xspace}
\newcommand{\gevcc}{\ensuremath{{\mathrm{\,Ge\kern -0.1em V\!/}c^2}}\xspace}
\newcommand{\gevgevcccc}{\ensuremath{{\mathrm{\,Ge\kern -0.1em V^2\!/}c^4}}\xspace}
\newcommand{\mevcc}{\ensuremath{{\mathrm{\,Me\kern -0.1em V\!/}c^2}}\xspace}

\def\m    {\ensuremath{\mathrm{ \,m}}\xspace}

\def\invfb   {\ensuremath{\mbox{\,fb}^{-1}}\xspace}

\def\gsim{{~\raise.15em\hbox{$>$}\kern-.85em
          \lower.35em\hbox{$\sim$}~}\xspace}
\def\lsim{{~\raise.15em\hbox{$<$}\kern-.85em
          \lower.35em\hbox{$\sim$}~}\xspace}

\newcommand{\Real}{\ensuremath{\mathcal{R}e}\xspace}
\newcommand{\Imag}{\ensuremath{\mathcal{I}m}\xspace}

\def\tell1  {TELL1\xspace}
\def\ukl1   {UKL1\xspace}

\newcommand{\eg}{\mbox{\itshape e.g.}\xspace}

\usepackage{cite} 
\usepackage{mciteplus}

\def\DzorDzbar                {\kern 0.18em\optbar{\kern -0.18em \Dz}{}\xspace}

\def\Y                        {\ensuremath{\mathcal{N}}\xspace}

\ifthenelse{\boolean{uprightparticles}}%
{
 
}
{
 
}

\def\Bdsbz      {\ensuremath{\Bbar^0_{(s)}}\xspace}

\def\BdorBdbar    {\ensuremath{\kern 0.18em\optbar{\kern -0.18em B}{}^0}\xspace}
\def\BsorBsbar    {\ensuremath{\kern 0.18em\optbar{\kern  0.06em B_s}{}^0}\xspace}

\def\BdtoKsKK   {\decay{\Bd}{\KS \Kp \Km}}
\def\BdtoKsPiPi   {\decay{\Bd}{\KS \pip \pim}}

\def\BdtoKsKpPim   {\decay{\Bd}{\KS \Kp \pim}}
\def\BdtoKsPipKm   {\decay{\Bd}{\KS \Km \pip}}

\def\BdbtoKsPiPi   {\decay{\Bdb}{\KS \pip \pim}}
\def\BdsbtoKsPiPi   {\decay{\Bdsbz}{\KS \pip \pim}}

\def\BstoKsKK   {\decay{\Bs}{\KS \Kp \Km}}
\def\BstoKsPiPi   {\decay{\Bs}{\KS \pip \pim}}
\def\BstoKsKPi   {\decay{\Bs}{\KS \Kpm \pimp}}
\def\BstoKsKpPim   {\decay{\Bs}{\KS \Kp \pim}}
\def\BstoKsPipKm   {\decay{\Bs}{\KS \Km \pip}}

\def\KsPiPi{\ensuremath{\KS \pip \pim}\xspace}

\def\KsKK{\ensuremath{\KS \Kp \Km}\xspace}

\def\KsKpPim{\ensuremath{\KS \Kp \pim}\xspace}
\def\KsPipKm{\ensuremath{\KS \pip \Km}\xspace}

\newcommand{\KSpipi}{\KS\pip\pim}

\def\BdbtoetapKs {\decay{\Bdb}{\etapr \KS}}

\def\btoqqbars {\decay{\bquark}{\quark\quarkbar\squark}}

\def\btoccbars {\decay{\bquark}{\cquark\cquarkbar\squark}}

\def\LL   {Long-Long}
\def\DD   {Down-Down}

\def \fitCombShape{Polynomial}
\def \fitCombModel{from5150}

\newcommand{\plotIfExists}[2]{
  \IfFileExists{#1}{\includegraphics[width=#2\textwidth]{#1}}{\includegraphics[width=#2\textwidth]{#1}}
}    
\newcommand{\plotOne}[2]{
  \ifthenelse{\boolean{pdflatex}}{
    \plotIfExists{#1.pdf}{#2}
  }{
    \plotIfExists{#1.eps}{#2}
  }
}

\newcommand{\plotDataFitResults}[3]{ 
  \begin{figure}[p]
    \begin{center}
      \ifthenelse{\equal{\fitCombShape}{Exponential}}
                 {\def \tempFitCombShape{Exponential}}
                 {\def \tempFitCombShape{Polynomial}}
                 
                 \plotOne{figs/FitResults/#1/\fitCombModel-Louis-\tempFitCombShape-StrongPcut-#1-KSKK#2_#3-Standard-DoubleCB}{0.35}
                 \plotOne{figs/FitResults/#1/\fitCombModel-Louis-\tempFitCombShape-StrongPcut-#1-KSKK#2_#3-Standard-DoubleCB_log}{0.35}\\
                 \plotOne{figs/FitResults/#1/\fitCombModel-Louis-\tempFitCombShape-StrongPcut-#1-KSKpi#2_#3-Standard-DoubleCB}{0.35}
                 \plotOne{figs/FitResults/#1/\fitCombModel-Louis-\tempFitCombShape-StrongPcut-#1-KSKpi#2_#3-Standard-DoubleCB_log}{0.35}\\
                 \plotOne{figs/FitResults/#1/\fitCombModel-Louis-\tempFitCombShape-StrongPcut-#1-KSpiK#2_#3-Standard-DoubleCB}{0.35}
                 \plotOne{figs/FitResults/#1/\fitCombModel-Louis-\tempFitCombShape-StrongPcut-#1-KSpiK#2_#3-Standard-DoubleCB_log}{0.35}\\
                 \plotOne{figs/FitResults/#1/\fitCombModel-Louis-\tempFitCombShape-StrongPcut-#1-KSpipi#2_#3-Standard-DoubleCB}{0.35}
                 \plotOne{figs/FitResults/#1/\fitCombModel-Louis-\tempFitCombShape-StrongPcut-#1-KSpipi#2_#3-Standard-DoubleCB_log}{0.35}\\
                 \ifthenelse{\equal{#2}{DD}}{
                   \caption{Results of the simultaneous fit to data (\DD, #3) with the \MakeLowercase{#1} BDT optimisation. The modes \KsKK, \KsKpPim, \KsPipKm and \KsPiPi are shown from top to bottom. The left-hand side plots show the results with a linear scale and the right-hand side with a logarithmic scale.}
                 }{
                   \caption{Results of the simultaneous fit to data (\LL, #3) with the \MakeLowercase{#1} BDT optimisation. The modes \KsKK, \KsKpPim, \KsPipKm and \KsPiPi are shown from top to bottom. The left-hand side plots show the results with a linear scale and the right-hand side with a logarithmic scale.}
                 }
                 \label{fig:FitResult:#1:#2:#3}
    \end{center}
  \end{figure}
}
\newcommand{\plotSignalMCFitResults}[3]{
  \begin{figure}[!htbp]
    \begin{center}
      \plotOne{figs/FitModel/Signal/#1/Bd2KSKK#2_#3-MCFit-Louis-DoubleCB-Standard_log}{0.35}
      \plotOne{figs/FitModel/Signal/#1/Bs2KSKK#2_#3-MCFit-Louis-DoubleCB-Standard_log}{0.35}\\
      \plotOne{figs/FitModel/Signal/#1/Bd2KSKpi#2_#3-MCFit-Louis-DoubleCB-Standard_log}{0.35}
      \plotOne{figs/FitModel/Signal/#1/Bs2KSKpi#2_#3-MCFit-Louis-DoubleCB-Standard_log}{0.35}\\
      \plotOne{figs/FitModel/Signal/#1/Bd2KSpiK#2_#3-MCFit-Louis-DoubleCB-Standard_log}{0.35}
      \plotOne{figs/FitModel/Signal/#1/Bs2KSpiK#2_#3-MCFit-Louis-DoubleCB-Standard_log}{0.35}\\
      \plotOne{figs/FitModel/Signal/#1/Bd2KSpipi#2_#3-MCFit-Louis-DoubleCB-Standard_log}{0.35}
      \plotOne{figs/FitModel/Signal/#1/Bs2KSpipi#2_#3-MCFit-Louis-DoubleCB-Standard_log}{0.35}\\
      \ifthenelse{\equal{#2}{DD}}{
        \caption{Result of the simultaneous fit to simulated samples of the signal decays for \DD \KS reconstruction mode, using the \MakeLowercase{#1} optimisation of the BDT, and shown using a logarithmic scale. \KsKK, \KsKpPim, \KsPipKm, and \KsPiPi are shown from top to bottom, while \Bd decays are shown on the left and \Bs decays on the right. }
      }
                 {
                   \caption{Result of the simultaneous fit to simulated samples of the signal decays for \LL \KS reconstruction mode, using the \MakeLowercase{#1} optimisation of the BDT, and shown using a logarithmic scale. \KsKK, \KsKpPim, \KsPipKm, and \KsPiPi are shown from top to bottom, while \Bd decays are shown on the left and \Bs decays on the right. }
                 }
                 \label{fig:FitModel:Signal:#1:#2:#3}
    \end{center}
\end{figure}
}

\newcommand{\plotSplots}[3]{
  \ifthenelse{\equal{#3}{log}}
             {\def\suffix{FitStandard_log}}
             {\def\suffix{FitStandard}}
             \ifthenelse{\equal{#3}{log}}
                        {\def\scale{logarithmic}\xspace}
                        {\def\scale{linear}\xspace}
                        \begin{figure}[!htbp]
                          \begin{center}
                            \plotOne{figs/FitResults/sWeights/#1/sWeights-from5150-Louis-PolSlopes-StrongPcut-#1-KSKKDD_#2_\suffix}{0.35}
                            \plotOne{figs/FitResults/sWeights/#1/sWeights-from5150-Louis-PolSlopes-StrongPcut-#1-KSKKLL_#2_\suffix}{0.35}\\
                            \plotOne{figs/FitResults/sWeights/#1/sWeights-from5150-Louis-PolSlopes-StrongPcut-#1-KSKpiDD_#2_\suffix}{0.35}
                            \plotOne{figs/FitResults/sWeights/#1/sWeights-from5150-Louis-PolSlopes-StrongPcut-#1-KSKpiLL_#2_\suffix}{0.35}\\
                            \plotOne{figs/FitResults/sWeights/#1/sWeights-from5150-Louis-PolSlopes-StrongPcut-#1-KSpiKDD_#2_\suffix}{0.35}
                            \plotOne{figs/FitResults/sWeights/#1/sWeights-from5150-Louis-PolSlopes-StrongPcut-#1-KSpiKLL_#2_\suffix}{0.35}\\
                            \plotOne{figs/FitResults/sWeights/#1/sWeights-from5150-Louis-PolSlopes-StrongPcut-#1-KSpipiDD_#2_\suffix}{0.35}
                            \plotOne{figs/FitResults/sWeights/#1/sWeights-from5150-Louis-PolSlopes-StrongPcut-#1-KSpipiLL_#2_\suffix}{0.35}\\
                            \caption{Result of the invariant mass fits used for the sWeights extraction with the \MakeLowercase{#1} BDT optimisation on #2 data (\scale~scale). \KsKK, \KsKpPim, \KsPipKm, and \KsPiPi are shown from top to bottom, \DD on the left, \LL on the right.}
                            \label{fig:FitResult:Splots:#1:#2:#3}
                          \end{center}
                        \end{figure}
}
\newcommand{\plotSplotsDalitz}[3]{ 
  \ifthenelse{\equal{\fitCombShape}{Exponential}}
             {\def \tempFitCombShape{Exponential}}
             {\def \tempFitCombShape{PolSlopes}}
             
             \begin{figure}[!htbp]
               \begin{center}
                 \plotOne{figs/FitResults/sWeights/#1/sWeights-\fitCombModel-Louis-\tempFitCombShape-StrongPcut-#1-KSKKDD_#3_#2_Dalitz}{0.35}
                 \plotOne{figs/FitResults/sWeights/#1/sWeights-\fitCombModel-Louis-\tempFitCombShape-StrongPcut-#1-KSKKLL_#3_#2_Dalitz}{0.35}\\
                 \plotOne{figs/FitResults/sWeights/#1/sWeights-\fitCombModel-Louis-\tempFitCombShape-StrongPcut-#1-KSKpiDD_#3_#2_Dalitz}{0.35}
                 \plotOne{figs/FitResults/sWeights/#1/sWeights-\fitCombModel-Louis-\tempFitCombShape-StrongPcut-#1-KSKpiLL_#3_#2_Dalitz}{0.35}\\
                 \plotOne{figs/FitResults/sWeights/#1/sWeights-\fitCombModel-Louis-\tempFitCombShape-StrongPcut-#1-KSpiKDD_#3_#2_Dalitz}{0.35}
                 \plotOne{figs/FitResults/sWeights/#1/sWeights-\fitCombModel-Louis-\tempFitCombShape-StrongPcut-#1-KSpiKLL_#3_#2_Dalitz}{0.35}\\
                 \plotOne{figs/FitResults/sWeights/#1/sWeights-\fitCombModel-Louis-\tempFitCombShape-StrongPcut-#1-KSpipiDD_#3_#2_Dalitz}{0.35}
                 \plotOne{figs/FitResults/sWeights/#1/sWeights-\fitCombModel-Louis-\tempFitCombShape-StrongPcut-#1-KSpipiLL_#3_#2_Dalitz}{0.35}\\
                 \ifthenelse{\equal{#2}{Bd}}{
                   \caption{Distribution of \Bd signal sWeights extracted with the \MakeLowercase{#1} BDT optimisation in #3 data. \KsKpPim, \KsPipKm, and \KsPiPi are shown from top to bottom, \DD on the left, \LL on the right. The corrections due to the presence of species with fixed yields is not applied here.}
                 }{
                   \caption{Distribution of \Bs signal sWeights extracted with the \MakeLowercase{#1} BDT optimisation in #3 data. \KsKpPim, \KsPipKm, and \KsPiPi are shown from top to bottom, \DD on the left, \LL on the right. The corrections due to the presence of species with fixed yields is not applied here.}
                 } 
                 \label{fig:FitResult:Splots:#1:#3:Dalitz#2}
               \end{center}
             \end{figure}
}

\newcommand{\makeSystTabular}[2]{
  \begin{table}[!htbp]
    \begin{center}
      \ifthenelse{\equal{#2}{KK}}
                 {       
                   \caption{Systematic uncertainties originating from the fit model of the \KsKK modes (using \MakeLowercase{#1} BDT optimisation). The numbers are rounded to the upper integer value, except for the total yield. The total systematic error is defined as the sum in quadrature of all the components.}
                 }{}
                 \ifthenelse{\equal{#2}{Kpi}}
                            {
                              \caption{Systematic uncertainties originating from the fit model of the \KsKpPim modes (using \MakeLowercase{#1} BDT optimisation). The numbers are rounded to the upper integer value, except for the total yield. The total systematic error is defined as the sum in quadrature of all the components.}
                            }{}
                            \ifthenelse{\equal{#2}{piK}}        
                                       {
                                         \caption{Systematic uncertainties originating from the fit model of the \KsPipKm modes (using \MakeLowercase{#1} BDT optimisation). The numbers are rounded to the upper integer value, except for the total yield. The total systematic error is defined as the sum in quadrature of all the components.}
                                       }{}
                                       \ifthenelse{\equal{#2}{pipi}}
                                                  {
                                                    \caption{Systematic uncertainties originating from the fit model of the \KsPiPi modes (using \MakeLowercase{#1} BDT optimisation). The numbers are rounded to the upper integer value, except for the total yield. The total systematic error is defined as the sum in quadrature of all the components.}
                                                  }{}
                                                  \label{Table:FitSyst:#1:#2}
                                                  \resizebox{\textwidth}{!}{
                                                    \input{tables/SystI-PolSlopes-from5150-#1-#2.txt}
                                                  }
    \end{center}
  \end{table}
  
}

\newcommand{\makeInvMass}[1]{
  \ifthenelse{\equal{#1}{Bd2KSpipi} \or \equal{#1}{Bs2KSpipi}}              {\def\myInvMass {pipi}}
             {\ifthenelse{\equal{#1}{Bd2KSpiK} \or \equal{#1}{Bs2KSpiK}}    {\def\myInvMass {piK}}
               {\ifthenelse{\equal{#1}{Bd2KSKpi} \or \equal{#1}{Bs2KSKpi}}  {\def\myInvMass {Kpi}}
                 {\ifthenelse{\equal{#1}{Bd2KSKK} \or \equal{#1}{Bs2KSKK}}  {\def\myInvMass {KK}}{\def\myInvMass ERROR}}
               }
             }

}
\newcommand{\makeMode}[1]{
  \ifthenelse{\equal{#1}{Bd2KSpipi}}                     {\def\myMode {\BdtoKsPiPi}}
             {\ifthenelse{\equal{#1}{Bd2KSpiK}}          {\def\myMode {\BdtoKsPipKm}}
               {\ifthenelse{\equal{#1}{Bd2KSKpi}}        {\def\myMode {\BdtoKsKpPim}}
                 {\ifthenelse{\equal{#1}{Bd2KSKK}}       {\def\myMode {\BdtoKsKK}}
                   {\ifthenelse{\equal{#1}{Bs2KSpipi}}           {\def\myMode {\BstoKsPiPi}}
                     {\ifthenelse{\equal{#1}{Bs2KSpiK}}          {\def\myMode {\BstoKsPipKm}}
                       {\ifthenelse{\equal{#1}{Bs2KSKpi}}        {\def\myMode {\BstoKsKpPim}}
                         {\ifthenelse{\equal{#1}{Bs2KSKK}}       {\def\myMode {\BstoKsKK}}{\def\myMode{ERROR}}
                         }
                       }
                     }
                   }
                 }
               }
             }
}

\newcommand{\plotGeomEff}[3]{
  \makeInvMass{#2}
  \makeMode{#2}
  \ifthenelse{\equal{#1}{2011}}{\def\redYear {2011}}{\def\redYear {2012}}
  \centering
  \includegraphics[width=0.45\textwidth]{figs/EfficiencyMaps/#1/Geometry/NoSel/#2/Efficiency_Map_NoSel_#2_#3_#1_\myInvMass.pdf.eps}
  \includegraphics[width=0.45\textwidth]{figs/EfficiencyMaps/#1/Geometry/NoSel/#2/Spline_Eff_NoSel_#2_#3_#1_\myInvMass.pdf.eps}
  \includegraphics[width=0.45\textwidth]{figs/EfficiencyMaps/#1/Geometry/NoSel/#2/Efficiency_errorHi_Map_NoSel_#2_#3_#1_\myInvMass.pdf.eps}
  \includegraphics[width=0.45\textwidth]{figs/EfficiencyMaps/#1/Geometry/NoSel/#2/Efficiency_errorLo_Map_NoSel_#2_#3_#1_\myInvMass.pdf.eps}
  \caption{
    (Top) $\epsilon^{\rm geom}$ as a function of the \myMode
    square Dalitz plot position obtained from \redYear-conditions generator-level signal MC:
    (left) the raw histogram,
    (right) smoothed using a 2D cubic spline.
    (Bottom) the (left) upper and (right) lower uncertainties on the histogram bins.
    Uncertainties are due to MC statistics.
  }
  \label{fig:geoeff:GeoEff:#1:#2:#3}
}

\newcommand{\plotTrackCorr}[4]{
  \makeInvMass{#2}
  \makeMode{#2}
  \centering
  \includegraphics[width=0.45\textwidth]{figs/EfficiencyMaps/#1/Tracking/#4/#2/Trk_AllEff-#4-#2-#3-#1-\myInvMass.pdf.eps}
  \includegraphics[width=0.45\textwidth]{figs/EfficiencyMaps/#1/Tracking/#4/#2/Spline_Eff_#4_#2_#3_#1_\myInvMass.pdf.eps}\\
  \includegraphics[width=0.45\textwidth]{figs/EfficiencyMaps/#1/Tracking/#4/#2/Tracking_errorHi_bootstrap_#4_#2_#3_\myInvMass_#1.pdf.eps}
  \includegraphics[width=0.45\textwidth]{figs/EfficiencyMaps/#1/Tracking/#4/#2/Tracking_errorHi_bootstrap_#4_#2_#3_\myInvMass_#1.pdf.eps}
  \caption{
    (Top) Combined tracking efficiency corrections in the \myMode signal mode for #3 and #1 configuration:
    (left) the raw histogram obtained from MC simulation and (right) smoothed using a 2D cubic spline.  
    (Bottom) the (left) upper and (right) lower uncertainties on the histogram bins.
  }
  \label{fig:DPefficiency-trk-all-#1-#2-#3-#4}
}

\newcommand{\plotTrackCorrMode}[1]{
  \begin{figure}[!htb]
    \plotTrackCorr{2011}{#1}{DD}{Loose}
  \end{figure}
  \begin{figure}[!htb]
    \plotTrackCorr{2011}{#1}{LL}{Loose}
  \end{figure}
}

\newcommand{\plotTrigCorr}[4]{
  \makeInvMass{#1}
  \makeMode{#1}
  \ifthenelse{\equal{#4}{TOS}}{
    \def\trigComment {$\epsilon^{\rm L0TOS|sel\&geom}_{\rm data}/\epsilon^{\rm L0TOS|sel\&geom}_{\rm MC}$}}{
    \def\trigComment {$\epsilon^{\rm !L0TOS|sel\&geom}_{\rm data}/\epsilon^{\rm !L0TOS|sel\&geom}_{\rm MC}$}}
  
  \centering
  \includegraphics[width=0.45\textwidth]{figs/EfficiencyMaps/2011/L0#4/#3/#1/L0#4-Correction-#3-#1-#2-2011-\myInvMass.pdf.eps}
  \includegraphics[width=0.45\textwidth]{figs/EfficiencyMaps/2011/L0#4/#3/#1/Spline_Eff_#3_#1_#2_2011_\myInvMass.pdf.eps}\\
  \includegraphics[width=0.45\textwidth]{figs/EfficiencyMaps/2011/L0#4/#3/#1/L0#4_errorHi_combined_#3_#1_#2_\myInvMass_2011.pdf.eps}
  \includegraphics[width=0.45\textwidth]{figs/EfficiencyMaps/2011/L0#4/#3/#1/L0#4_errorLo_combined_#3_#1_#2_\myInvMass_2011.pdf.eps}
  \caption{
    (Top) \trigComment across the \myMode #2 square Dalitz plot (2011+2012 combined):
    (left) the raw histogram obtained using the procedure described in the text and (right) smoothed using a 2D cubic spline.  
    (Bottom) the (left) upper and (right) lower uncertainties on the histogram bins.
  }
  \label{fig:DPcorrection-#1-#2-#3-#4}
}

\newcommand{\plotSelEff}[5]{
  \makeInvMass{#2}
  \makeMode{#2}
  \ifthenelse{\equal{#5}{TOS}}{\def\selComment {{\tt L0Hadron\_TOS}}}{\def\selComment {{\tt L0Global\_TIS\&\&!L0Hadron\_TOS}}}
  \centering
  \includegraphics[width=0.45\textwidth]{figs/EfficiencyMaps/#1/Sel#5/#4/#2/#2-Eff-#4-#2-#3-#1-\myInvMass.pdf.eps}
  \includegraphics[width=0.45\textwidth]{figs/EfficiencyMaps/#1/Sel#5/#4/#2/Spline_Eff_#4_#2_#3_#1_\myInvMass.pdf.eps}\\
  \includegraphics[width=0.45\textwidth]{figs/EfficiencyMaps/#1/Sel#5/#4/#2/#2-ErrorHi-#4-#2-#3-#1-\myInvMass.pdf.eps}
  \includegraphics[width=0.45\textwidth]{figs/EfficiencyMaps/#1/Sel#5/#4/#2/#2-ErrorLo-#4-#2-#3-#1-\myInvMass.pdf.eps}
  \caption{
    (Top) $\epsilon^{\rm sel|geom}$ across the \myMode #3 #1 square Dalitz plot for \selComment \MakeLowercase{#4} BDT candidates:
    (left) the raw histogram obtained using the procedure described in the text and (right) smoothed using a 2D cubic spline.  
    (Bottom) the (left) upper and (right) lower uncertainties on the histogram bins.
  }
  \label{fig:DPefficiency-sel-#1-#2-#3-#4-#5}
}

\newcommand{\plotSelEffMode}[2]{
  \begin{figure}[!htb]
    \plotSelEff{2011}{#1}{DD}{#1}{TOS}
  \end{figure}
  \begin{figure}[!htb]
    \plotSelEff{2011}{#1}{DD}{#1}{TIS}
  \end{figure}
  \begin{figure}[!htb]
    \plotSelEff{2011}{#1}{LL}{#1}{TOS}
  \end{figure}
  \begin{figure}[!htb]
    \plotSelEff{2011}{#1}{LL}{#1}{TIS}
  \end{figure}
}

\usepackage{longtable} 
\usepackage{multirow}
\usepackage{gensymb}
\usepackage{placeins}

\newcommand{\specialcell}[2][c]{ %
  \begin{tabular}[#1]{@{}c@{}}#2\end{tabular}}

\begin{document}
\renewcommand{\thefootnote}{\fnsymbol{footnote}}
\setcounter{footnote}{1}


\begin{titlepage}
\pagenumbering{roman}

\vspace*{-1.5cm}
\centerline{\large EUROPEAN ORGANIZATION FOR NUCLEAR RESEARCH (CERN)}
\vspace*{1.5cm}
\noindent
\begin{tabular*}{\linewidth}{lc@{\extracolsep{\fill}}r@{\extracolsep{0pt}}}
\ifthenelse{\boolean{pdflatex}}
{\vspace*{-1.5cm}\mbox{\!\!\!\includegraphics[width=.14\textwidth]{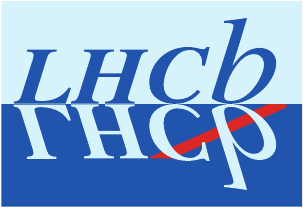}} & &}%
{\vspace*{-1.2cm}\mbox{\!\!\!\includegraphics[width=.12\textwidth]{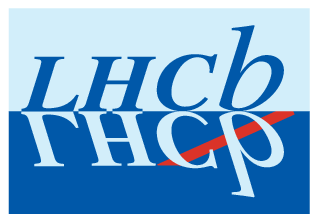}} & &}%
\\
 & & CERN-EP-2017-317 \\  
 & & LHCb-PAPER-2017-033 \\  
 & & 22 December 2017\\ 
\end{tabular*}

\vspace*{4.0cm}

{\normalfont\bfseries\boldmath\huge
\begin{center}
  \papertitle 
\end{center}
}

\vspace*{1.5 cm}

\begin{center}
\paperauthors\footnote{Authors are listed at the end of this paper.}
\end{center}

\vspace{\fill}

\begin{abstract}
   
   \noindent
  The time-integrated Dalitz plot of the three-body hadronic charmless decay ${\Bdb \to K_{\mathrm{\scriptscriptstyle S}}^0 \pi^+ \pi^-}$  is studied using a $pp$ collision data sample recorded with the \lhcb detector,  corresponding to an integrated luminosity of $3.0\mbox{\,fb}^{-1}$. The decay amplitude is described with an isobar model. Relative contributions of the isobar amplitudes to the ${\Bdb \to K_{\mathrm{\scriptscriptstyle S}}^0 \pi^+ \pi^-}$ decay branching fraction and \CP asymmetries of the flavour-specific amplitudes are measured. The \CP asymmetry between the conjugate ${\Bdb  \to K^{*}(892)^{-}\pi^+}$   and  ${\Bd \to K^{*}(892)^{+}\pi^-}$ decay rates is determined to be 
 $-0.308 \pm 0.062$.     
\end{abstract}

\vspace*{2.0cm}

\begin{center}
  Submitted to Phys. Rev. Lett. 
\end{center}

\vspace{\fill}

{\footnotesize 
\centerline{\copyright~\papercopyright, licence \href{\paperlicenceurl}{\paperlicence}.}}
\vspace*{2mm}

\end{titlepage}


\newpage
\setcounter{page}{2}
\mbox{~}
%
%
%
%

\cleardoublepage


\renewcommand{\thefootnote}{\arabic{footnote}}
\setcounter{footnote}{0}



\pagestyle{plain} 
\setcounter{page}{1}
\pagenumbering{arabic}


The breaking of the invariance of the weak interaction under the combined action of the charge conjugation ($C$) and parity ($P$)  transformations 
is firmly established in the $K$- and $B$-meson systems~\cite{Christenson:1964fg,Aubert:2001nu,Abe:2001xe}. In particular, significant  \CP 
asymmetries at the level of 10\% or more have been measured in the decays of $B$ mesons into two light pseudoscalars. The \CP asymmetries 
in the decays of  $\Bdb \to K^- \pi^+$ and $B^- \to K^- \pi^0$ are observed to be different~\cite{HFAG}, while, in predictions based on the QCD factorisation 
approach, the two asymmetries are expected to be similar~\cite{Beneke:2003zv}.  This apparent discrepancy is often referred to in the literature as the 
$K\pi$ puzzle~\cite{Fleischer:2007mq,Baek:2009pa,Li:2009wba,Khalil:2009zf}. The study of the flavour-specific quasi-two-body amplitudes which contribute 
to the decay  \BdbtoKsPiPi offers the possibility to measure \CP asymmetries. In particular, the decays with a vector and a pseudoscalar in the final state, 
such as  ${\overline{B}}^0 \to K^{*}(892)^- \pi^+$, may help to shed light on the $K\pi$ puzzle.   
\par
In the Standard Model (SM)~\cite{Cabibbo:1963yz,Kobayashi:1973fv}, the mixing-induced \CP asymmetries in the quark-level transitions  \btoqqbars 
($q = u,d,s$) which govern the decay \BdbtoKsPiPi are predicted to be approximately equal to those in  \btoccbars transitions, 
such as $\Bd\to\jpsi\KS$. The existence of new particles in extensions of the SM could introduce additional weak phases that 
contribute along with the SM mixing phase~\cite{Buchalla:2005us,Grossman:1996ke,London:1997zk,Ciuchini:1997zp}. In general, for each of 
the studied \CP eigenstates, the current  experimental measurements of \btoqqbars decays~\cite{HFAG} show good agreement with 
the results from \btoccbars  decays~\cite{HFAG}.  There is nonetheless room for contributions from physics beyond the SM and, hence, the need 
for precision measurements of these weak mixing phases.
\par 
The mixing-induced \CP-violating phase can be measured by means of a decay-time-dependent analysis of the Dalitz plot 
(DP)~\cite{Dalitz:1953cp} of the decay  \BdbtoKsPiPi ~\cite{Dalseno:2008wwa,Aubert:2009me,Nakahama:2010nj,Lees:2012kxa}.  Such an analysis 
requires the initial flavour of the \Bdb meson to be determined or ``tagged". A recent study of the yields of the charmless three-body decays \BdbtoKsPiPi 
has been reported in Ref.~\cite{Aaij:2017zpx}. The \BdbtoKsPiPi yields are comparable to those obtained at the \babar and \belle experiments 
but the lower tagging efficiency at \lhcb does not yet allow a precise flavour-tagged analysis to be performed. The decay-time-integrated untagged DP 
of this mode is studied in this Letter.  The DP of the decay \BdbtoKsPiPi is modelled by a sum of quasi-two-body amplitudes (the isobar parameterisation) 
and the model is fit to the \lhcb data to measure the relative branching fractions and the \CP 
asymmetries of flavour-specific final states.  
\par
The analysis reported in this Letter is performed using $pp$ collision data recorded with the LHCb detector, corresponding to integrated luminosities  
of 1.0\invfb at a centre-of-mass energy of 7\tev in 2011 and to 2.0\invfb at a centre-of-mass energy of  8\tev in 2012. The \lhcb detector~\cite{Alves:2008zz,LHCb-DP-2014-002} 
is a single-arm forward spectrometer covering the \mbox{pseudorapidity} range $2<\eta <5$, designed for the study of particles containing \bquark or 
\cquark quarks. Signal candidates are accepted if one of the final-state particles from the signal decay deposits sufficient energy transverse to the beamline in the hadronic 
calorimeter to pass the hardware trigger. Events that are triggered at the hardware level by another particle in the event are also retained.  In a second step, 
a software trigger requires a two-, three- or four-track secondary vertex with a significant displacement from any primary $pp$ interaction vertex (PV). At least one 
charged particle must have a large transverse momentum and be inconsistent with originating from a PV. A multivariate algorithm~\cite{BBDT} is 
used for the identification of secondary vertices consistent with the decay of a \bquark hadron.
\par
The selection procedure is described in detail in Ref.~\cite{Aaij:2017zpx}. Decays of \decay{\KS}{\pip\pim} are reconstructed in two different 
categories: the first involving \KS mesons that decay early enough for the resulting pions to be reconstructed in the vertex detector; and the second containing 
those \KS mesons that decay later, such that track segments of the pions cannot be formed in the vertex detector. These categories are referred to as 
Long and Downstream, respectively. Downstream \KS were not reconstructed in the software trigger in 2011, but they were reconstructed and 
used for triggering in 2012.  Furthermore, an improved software trigger with larger $b$-hadron  efficiency, in particular in the Downstream category, was used for the 
second part of the 2012 data taking.  To take into account the different levels of trigger efficiency, the data sample is divided into 2011, 2012a, and 2012b 
data-taking periods, and each period is further divided according to the \KS reconstruction category, giving a total of six subsamples. The 2012b sample is 
the largest, corresponding to  an integrated luminosity 1.4\invfb, and has the highest trigger efficiency. 
\par
\begin{figure}[t]
\begin{center}
\includegraphics[width=0.75\textwidth]{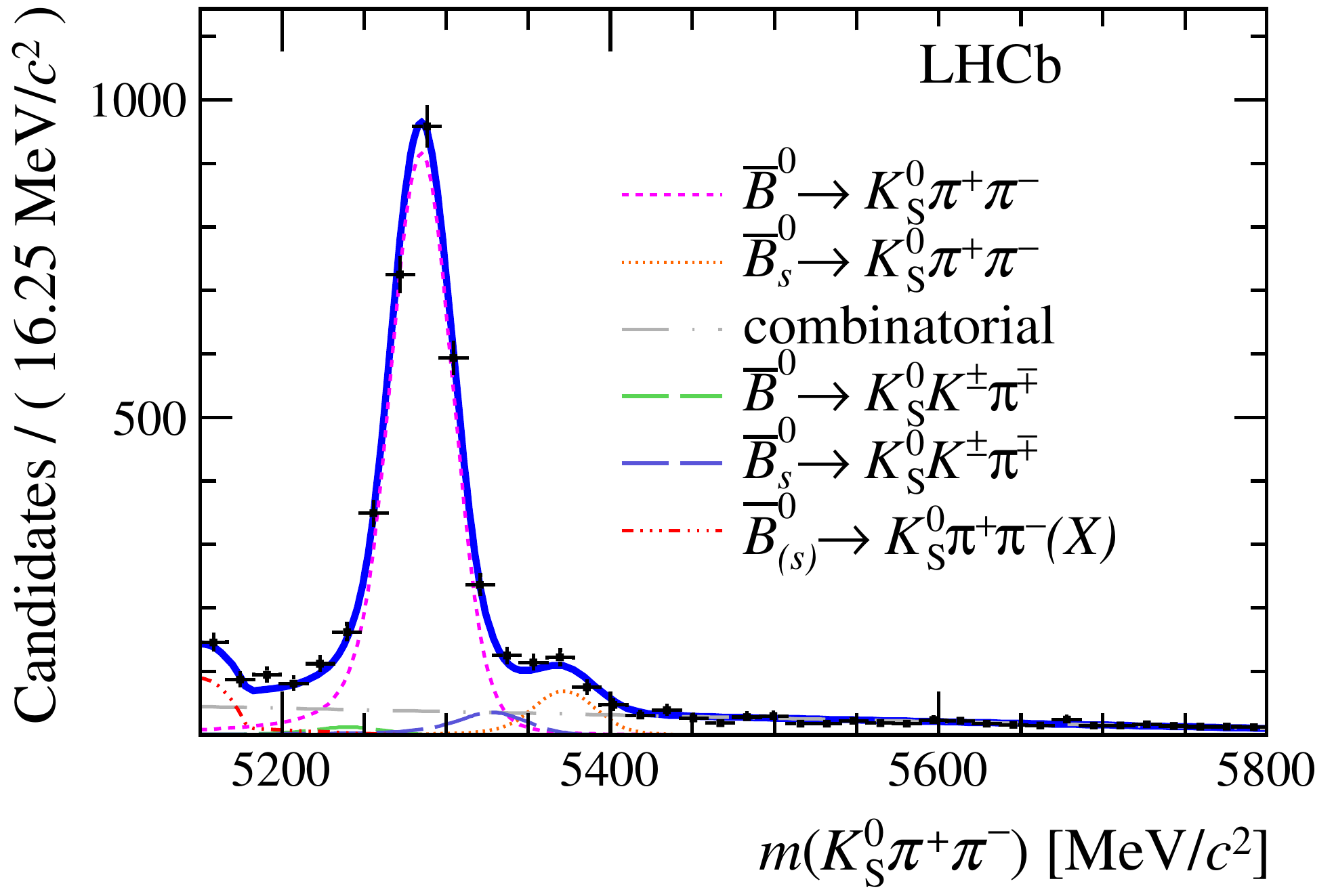}
\caption{Invariant mass distributions of \KsPiPi  candidates,  summing the two years of
                 data taking and the two \KS reconstruction categories. The sum of the partially 
                 reconstructed contributions from \B to open charm decays, charmless hadronic 
                 decays, \BdbtoetapKs and charmless radiative decays are denoted $\BdsbtoKsPiPi (X)$.}
\label{fig:fitTight}
\end{center}
\end{figure}

The events passing the trigger requirements are then filtered in two stages. Initial requirements are applied to further reduce the size of the data sample 
and increase the signal purity, before a multivariate classifier, based mostly on topological variables derived from the vertexing of the candidates, is 
implemented~\cite{Aaij:2017zpx}. The selection requirement placed on the output of the multivariate classifier is defined for each data subsample to 
yield a signal purity close to  90\%. Particle identification (PID) requirements are applied in order to reduce backgrounds from decays where either a proton, kaon or muon 
is misidentified as a pion. These criteria are optimised to reduce the cross-feed background coming from the decays \BstoKsKPi, where the kaon is misidentified 
as a pion.   The same invariant-mass vetoes on charmed and charmonium resonances as in Ref.~\cite{Aaij:2017zpx} are used in this analysis. The 
invariant-mass distribution of signal candidates from the six aforementioned subsamples is displayed in Fig.~\ref{fig:fitTight} with the result of a simultaneous fit. 
The candidates selected for the subsequent DP analysis are those in the \KSpipi mass range [5227,5343]~\mevcc.  

The DP analysis technique~\cite{Dalitz:1953cp} is employed to study the dynamics of the three-body decay \BdbtoKsPiPi.  
A decay-time-integrated untagged probability density function (p.d.f.)  is built to describe the phase space of the decay as a function of the DP kinematical variables. In this case, 
the p.d.f.  becomes an incoherent sum of  the $\mathcal{A}(s_+,s_-)$ and  $\kern 0.18em \overline{\kern -0.18em \mathcal{A}}(s_+,s_-)$  Lorentz-invariant transition amplitudes of the decays \BdtoKsPiPi  and  
 \BdbtoKsPiPi, respectively, 

\begin{equation}
\mathcal{P}(s_+,s_-) = \frac{|\mathcal{A}(s_+,s_-)|^2 + | \kern 0.18em \overline{ \kern -0.18em \mathcal{A}}(s_+,s_-)|^2}{\iint_{\rm DP} \left(|\mathcal{A}(s_+,s_-)|^2 + | \kern 0.18em \overline{ \kern -0.18em \mathcal{A}}(s_+,s_-)|^2\right ){\rm d}s_+{\rm d}s_-} \, ,\\
\label{eq:eqDP}
\end{equation} 

\noindent
where the kinematical variables $s_{\pm}$ denote the mass squared, $m_{\KS \pi^{\pm}}^2$.   
\par
The total amplitude $\overline{\mathcal{A}}(s_+,s_-)$  of the decay  \BdbtoKsPiPi is described as a coherent sum of the amplitudes of possible intermediate resonances and 
nonresonant contributions. The decay amplitudes for \Bd and \Bdb are given by   

\begin{equation}
\mathcal{A} = \sum_{j=1}^N c_jF_j(s_+,s_-) \, , \,\,
\overline{\mathcal{A}} = \sum_{j=1}^N \overline{c}_j \kern 0.18em \overline{\kern -0.18em F}_j(s_+,s_-) \, ,   
\end{equation}

\noindent
where $F_j$ and $\kern 0.18em \overline{\kern -0.18em F}_j$ are the DP spin-dependent dynamical functions for the resonance $j$ and $c_j$ are complex coefficients that account for the relative 
magnitudes and phases of the $N$ intermediate (resonant and nonresonant) components. The spin-dependent functions $F_j(s_+,s_-)$, embedding the resonance 
lineshape and the angular distributions, are constructed in the Zemach tensor formalism~\cite{Zemach:1965zz}. The weak-phase dependence is included in 
the $c_j$ coefficients. The results obtained for each isobar amplitude are expressed in this paper as a magnitude and a phase.  

The analysis method consists of a simultaneous DP fit to the six data subsamples defined above, with the shared isobar parameters determined using an unbinned maximum likelihood fit.  The DP model is built starting from the most significant amplitudes as determined in previous studies~\cite{Dalseno:2008wwa,Aubert:2009me,Nakahama:2010nj,Lees:2012kxa}. An algorithm to select the relevant additional amplitudes is defined before examining the data. A resonant amplitude is retained in the DP model if at least one of the following requirements is met:  (1) a goodness-of-fit estimator based on 
the point-to-point dissimilarity test~\cite{Williams:2010vh} decreases when the component is removed from the fit, (2) the likelihood ratio of the two hypotheses 
(component in and out) decreases, or (3) the significance of the magnitude of the component is at least three statistical standard deviations, neglecting systematic uncertainties. In particular, the 
components of the isobar DP model, $f_0(1500)\KS$ and  $K^{*}(1680)^{-}\pi^{+}$, which were not considered in previous studies, meet all three criteria. By contrast, 
the amplitude $f_2(1270)\KS$ is not retained. 

\begin{table}[!htbp]
\caption{
 Components of the DP model used in the fit.  The individual amplitudes are referred to by the resonance they contain.  
 The parameter values are given in \mevcc for the masses and \mev for the widths, except for $f_0(980)$ resonance. The parameter $m_0$ is the pole mass of the resonance and $\Gamma_0$ its natural width.  
 The mass-dependent lineshapes employed to model the resonances are indicated in the third column. Relativistic Breit-Wigner 
 and Gounaris-Sakurai lineshapes are denoted RBW and GS, respectively. EFKLLM is a parameterisation of the $\KS \pi^{-}$ S-wave lineshape, $(K\pi)^{-}_0$.  
}

\label{Tab: nominalmodel}
\begin{center}
\begin{tabular}{ c  c  c  c}
Resonance                & Parameters  & Lineshape  & Value references \\
\hline
$K^{*}(892)^{-}$        & \specialcell{$m_0 = 891.66 \pm 0.26$\\ $\Gamma_0 = 50.8 \pm 0.9$}  & RBW  & \cite{PDG2016}  \\\hline
$(K\pi)^{-}_0$         & \specialcell{$\Real(\lambda_0) = 0.204 \pm 0.103 $ \\ $\Imag(\lambda_0) = 0$ \\ $\Real(\lambda_1) = 1$ \\ $\Imag(\lambda_1) = 0$}  & EFKLLM \cite{ElBennich:2009da} & \cite{ElBennich:2009da}  \\\hline
$K^{*}_2(1430)^{-}$  & \specialcell{$m_0 = 1425.6 \pm 1.5$\\ $\Gamma_0 = 98.5 \pm 2.7$}  & RBW  & \cite{PDG2016}  \\\hline
$K^{*}(1680 )^{-}$  & \specialcell{$m_0 = 1717 \pm 27$\\ $\Gamma_0 = 332 \pm 110$}  & Flatt\'e ~\cite{Flatte:1976xu}  & \cite{PDG2016}  \\\hline
$f_0(500)$                 & \specialcell{$\m_0 = 513 \pm 32$ \\ $\Gamma_0 = 335 \pm 67$}   &  RBW  &  \cite{Muramatsu:2002jp} \\\hline
$\rho(770)^0$            & \specialcell{$m_0 = 775.26 \pm 0.25$ \\ $\Gamma_0 = 149.8 \pm 0.8$}  & GS ~\cite{Gounaris:1968mw} & \cite{PDG2016}  \\\hline
$f_0(980)$                 & \specialcell{$m_0 = 965 \pm 10$ \\ $g_{\pi} = 0.165 \pm 0.025$ GeV \\ $g_{K} = 0.695 \pm 0.119$ GeV}  & Flatt\'e  & \cite{Ablikim:2004wn}  \\\hline
$f_0(1500)$               & \specialcell{$ m_0 = 1505 \pm 6$ \\ $\Gamma_0 = 109 \pm 7$}         & RBW  &  \cite{PDG2016} \\\hline
$\chi_{c0}$                & \specialcell{$m_0 = 3414.75 \pm 0.31$ \\ $\Gamma_0 = 10.5 \pm 0.6$}  & RBW  & \cite{PDG2016}  \\\hline
Nonresonant (NR)  &                                          									&  Phase space & \\\hline
\end{tabular}
\end{center}
\end{table}

The signal DP model p.d.f.  $\mathcal{S}(s_+,s_-)$ is built from the coherent sum of the amplitudes listed in Table~\ref{Tab: nominalmodel}, normalising each isobar coefficient 
to the $K^{*}(892)^+\pi^-$ reference amplitude. The choice of the $K^{*}(892)^{\pm}\pi^{\mp}$ amplitudes as a reference provides the most stable DP fit. The phases of the reference 
amplitude and its conjugate are fixed to zero and the magnitude of the reference amplitude is arbitrarily fixed at 2.   
\par  
Two dominant backgrounds contaminate the \BdbtoKsPiPi candidate samples: a combinatorial background and a cross-feed background from the decay $\Bsb \to \KS K^{\pm}\pi^{\mp}$. 
The fractions of these backgrounds are measured from the invariant-mass fits performed in Ref.~\cite{Aaij:2017zpx} and their DP distributions are determined from the data. 
The combinatorial background DP model is built from the DP histogram of the \BdbtoKsPiPi  candidates with an invariant mass larger than 5450 \mevcc.  The DP model of the cross-feed 
background is measured from $\Bsb \to \KS K^{\pm}\pi^{\mp}$ candidates, where the $K^{\pm}$ is reconstructed under the $\pi^{\pm}$ hypothesis~\cite{LHCb-PAPER-2017-10}. 
The signal fraction depends on the reconstruction category; it is determined  from the fit to the invariant-mass distribution and ranges from 85\% (Downstream) to 95\% (Long). The p.d.f. in 
Eq.~\ref{eq:eqDP} is modified to account for the background components and the signal reconstruction efficiency across the DP, as determined from simulated events.  
\par
Two additional observables are formed from the isobar complex coefficients and are measured in the simultaneous DP fit.  The asymmetry observables $\mathcal{A}_{\rm raw}$ are derived 
from the measured isobar parameters of an amplitude $j$, $c_j$ and $\overline{c}_j$
\begin{eqnarray}
\mathcal{A}_{\rm raw} &=& \frac{|\overline{c}_j|^2 - |c_j|^2}{|\overline{c}_j|^2 + |c_j|^2}. 
\end{eqnarray}  
These observables are directly measured for flavour-specific final states. By contrast, the asymmetry of the mode $\overline{B}^0 \to f_0(980)\KS$ is determined using the patterns 
of its interference with flavour-specific amplitudes.  The \CP asymmetry is related to the raw asymmetry by $\mathcal{A}_{\CP} = \mathcal{A}_{\rm raw} - \mathcal{A}_{\Delta}$. The correction asymmetry is defined at first order as $\mathcal{A}_{\Delta} = A_P(\Bd) + A_D(\pi)$, where $A_P(\Bd)$ is the production asymmetry between the \Bd and  \Bdb mesons  and $A_D(\pi)$ is the detection asymmetry between \pip and \pim mesons.   
 The production asymmetry $A_P(\Bd)$ has been determined to be $A_P(\Bd) = (-0.35 \pm 0.81) \%$~\cite{LHCb-PAPER-2013-018}. Using $D_s^+$ decay modes~\cite{LHCb-PAPER-2012-009}, 
the pion detection asymmetry is measured to be consistent with zero with a 0.25\% uncertainty. The difference in the nuclear cross-sections for \Kz and \Kzb  
interactions in material results in a negligible bias~\cite{Ko:2010mk}. The uncertainty due to the correction asymmetries and the experimental systematic uncertainty are added in quadrature. 

The rate of a single process is proportional to the square of the relevant matrix element (see Eq.~\ref{eq:eqDP}). This involves the ensemble 
of its interferences with other components.  It is convenient to define the fit fraction of the process $i$, ${\cal F}_i$, as 

\begin{equation}
{\cal F}_i= \frac{\iint_{\rm DP} \left | c_i F_i(s_+,s_-) \right |^2 {\rm d}s_+{\rm d}s_-}{\iint_{\rm DP} \left | \sum_j c_j F_j(s_+,s_-) \right |^2{\rm d}s_+{\rm d}s_-}\;. 
\label{eq:FF}
\end{equation}  
 
Simulation is used to determine the selection efficiency of the signal. The simulation does not perfectly reproduce the detector response and these imperfections are corrected 
for in several respects. Firstly, the particle identification and misidentification efficiencies are determined from a calibration sample using reconstructed  $\Dstarp \to \Dz \pi^+$ decays, where 
the \Dz meson decays to the Cabibbo-favoured $\Km\pip$ final state. The variation of the PID performance with the track kinematics is included in the procedure. The calibration is performed using samples from the same 
data-taking period, accounting for the variation in the performance of the hadron identification detectors over time. 
Secondly, inaccuracies of the tracking simulation are mitigated by a weighting of the simulated tracking efficiency to match that measured in a calibration sample~\cite{LHCb-PUB-2011-025}. 
Analogous corrections are applied to the \KS decay-products tracking and vertexing efficiencies. Finally, a control sample of $\Dstarp \to \Dz (\to \Km\pip) \pi^+$ decays is used to quantify the differences of the 
hardware trigger response in data and simulation for pions and kaons, separated by positive and negative hadron charges, as a function of their transverse momentum~\cite{LHCb-DP-2012-004}.  The uncertainties 
assigned to these corrections are taken as a source of systematic uncertainties.

Two categories of systematic uncertainties are considered: experimental and related to the DP model. The former category comprises the uncertainties
on the fraction of signal, the fit biases,  the variation of the signal efficiency across the DP (including the choice of the efficiency binning) and the background DP models. The DP model  
uncertainties arise from the limited knowledge of the fixed parameters of the resonance-lineshape models, the marginal components neglected in the amplitude fit model and the modelling of the 
$\KS\pi^-$ and $\pi^+\pi^-$ S-wave components.
\par
All experimental uncertainties are estimated by means of pseudoexperiments,  in which samples for each reconstruction category are simulated and fitted exactly as for the data sample. For each 
pseudoexperiment, a single parameter governing a systematic effect (\eg the signal fraction) is varied according to its uncertainty. The standard deviation of the distribution of the fit results in an 
ensemble of 500 pseudoexperiments is taken as the corresponding systematic error estimate. 
The largest biases observed are at the few percent level.    
The final result is corrected for any observed bias where  it is significant. The dominant contribution to the experimental uncertainty is the efficiency determination.  
\par
The mass and the width of each resonance given in Table~\ref{Tab: nominalmodel} are varied individually and symmetrically by one standard deviation to evaluate the impact of the fixed parameters 
of the isobar resonance lineshapes.  The Blatt-Weisskopf radius parameter, fixed at $4 \gev^{-1}$, is varied by $\pm 1 \gev^{-1}$. 
\par
To evaluate the systematic uncertainties related to the marginal components of the DP model, the effect of adding the resonance $f_2(1270)$ (which is not retained by the previous criteria) and removing of the $f_0(500)$ component (the least significant contribution in the nominal model) is considered by repeating the fit with and without these components. 
Based upon this new model, a pseudoexperiment with a signal yield much larger than that of the data  is then generated and fit back with the nominal 
model. The related systematic uncertainty estimate is taken as the difference between the generated and fitted values. 
\par
A critical part of the isobar model design is the description of $K_S^0\pi^{\pm}$ S-wave components. Two parameterisations of these contributions have been studied: LASS~\cite{Aston:1987ir} and 
EFKLLM~\cite{ElBennich:2009da}.  The latter provides the best fit to the data. The log-likelihood difference between the two model hypotheses is $-2\Delta \ln {\cal L} = 85$. Given this large difference, no systematic uncertainty is then assigned to the choice of the EFKLLM parameterisation.  All model uncertainties are combined in quadrature to form the total model systematic uncertainty
\par
\begin{figure}[!htbp]
\begin{center}
\includegraphics[width=0.48\textwidth]{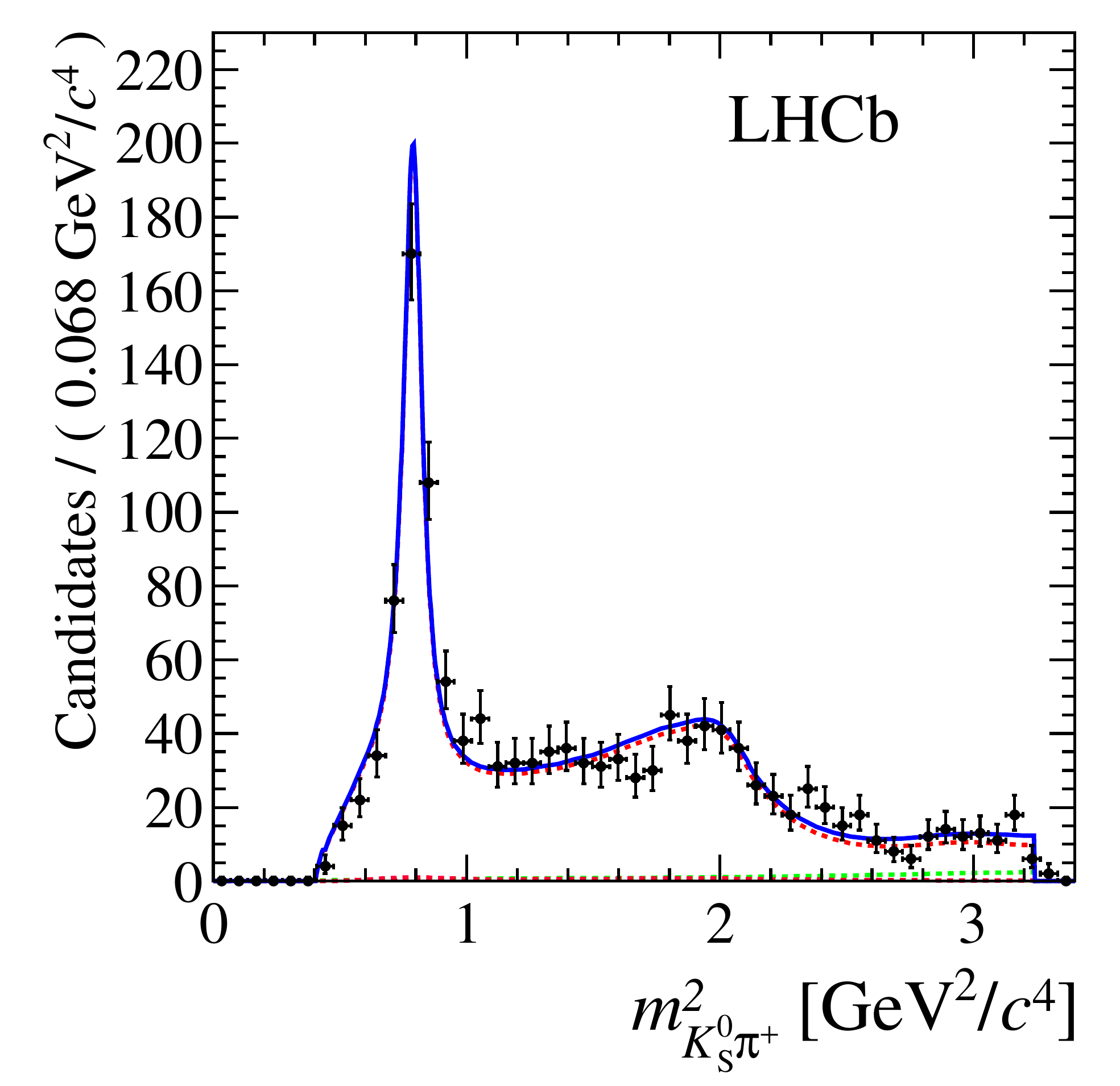}
\includegraphics[width=0.48\textwidth]{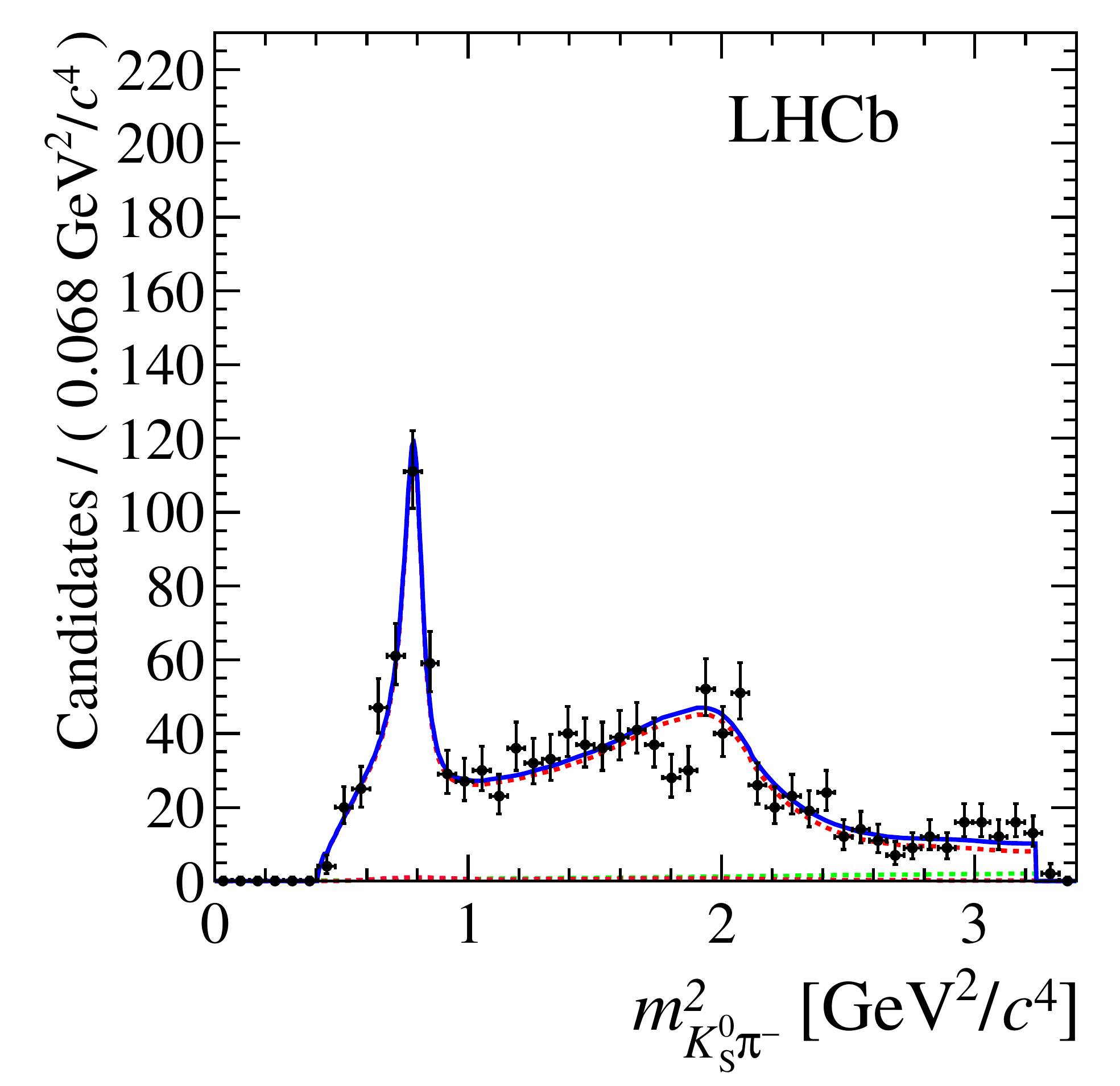}\\
\includegraphics[width=0.48\textwidth]{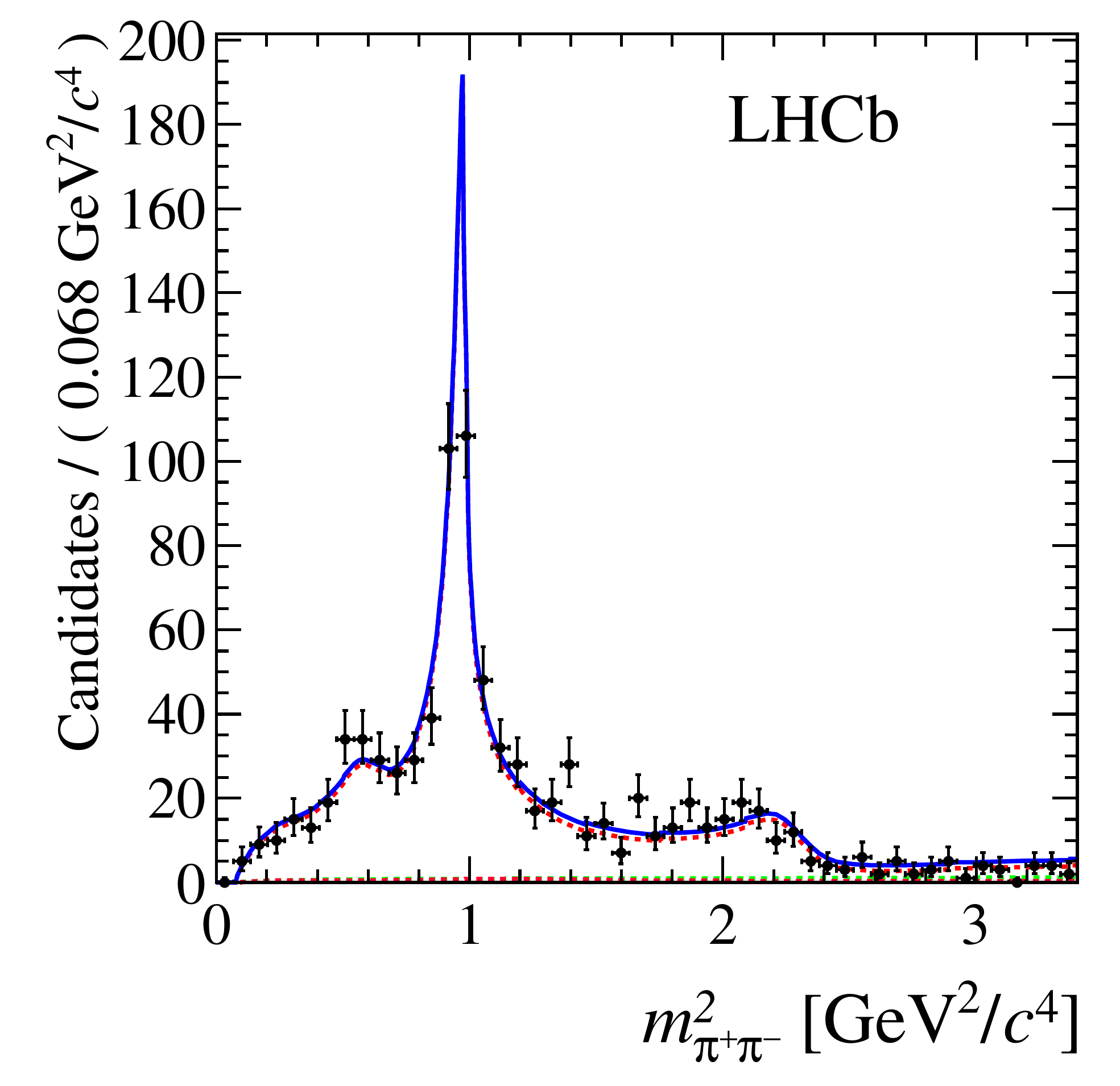}
\caption{  Projections of the sum of all data categories (black points) and the nominal fit function onto the DP variables (left) $m_{K_{\rm S}^0\pi^+}^2$, 
 (right) $m_{K_{\rm S}^0\pi^-}^2$ and (bottom) $m_{\pi^+\pi^-}^2$, restricted to the two-body low invariant-mass regions. The full fit is shown 
by the solid blue line and the signal model by the dashed red line. The observed difference is due to the (green) combinatorial and (light red) cross-feed background contributions,  
barely visible in these projections.}
\label{Fig : nominalfitzoom_all}
\end{center}
\end{figure}

The Dalitz plot projections are shown  in Fig.~\ref{Fig : nominalfitzoom_all} with the result of the fit superimposed.
The \CP-averaged fit fractions related to the quasi two-body and nonresonant amplitudes are derived from the isobar coefficients with Eq.~\ref{eq:FF}

\[\begin{aligned}
&\mathcal{F}(K^{*}(892)^{-}\pi^{+})      &=& \quad 9.43 \pm 0.40  \pm 0.33 \pm 0.34 \; \% \, , \nonumber \\
&\mathcal{F}((K\pi)^{-}_0\pi^{+})           &=& \quad 32.7 \pm  \,\,\, 1.4   \pm  \,\,\,  1.5   \pm  \,\,\, 1.1 \;   \%  \, ,  \nonumber \\
&\mathcal{F}(K^{*}_2(1430)^{-}\pi^{+}) &=& \quad 2.45 \,\, _{-\;\;\;\; 0.08}^{+\;\;\;\; 0.10}   \pm 0.14   \pm 0.12 \;  \%  \, , \nonumber \\
&\mathcal{F}(K^{*}(1680)^{-}\pi^{+})     &=& \quad 7.34 \pm 0.30   \pm  0.31 \pm 0.06 \;   \% \, , \nonumber \\
&\mathcal{F}(f_0(980)\KS)                               &=& \quad 18.6  \pm \,\,\,0.8  \pm \,\,\, 0.7 \pm \,\,\,1.2 \;   \%  \, , \nonumber \\
&\mathcal{F}(\rho(770)^0\KS)                         &=& \quad \,\,\, 3.8 \,\, _{-\;\;\;\;\;\;1.6}^{+\;\;\;\;\;\;1.1}  \pm \,\,\, 0.7   \pm \,\,\, 0.4 \;  \%  \, , \nonumber \\
&\mathcal{F}(f_0(500)\KS)                              &=& \quad 0.32 \,\, _{-\;\;\;\;0.08}^{+\;\;\;\;0.40}  \pm 0.19  \pm 0.23 \;   \% \, , \nonumber \\
&\mathcal{F}(f_0(1500)\KS)                            &=& \quad 2.60 \pm 0.54  \pm 1.28  \pm 0.60  \;  \% \;, \nonumber \\
&\mathcal{F}(\chi_{c0}\KS)                             &=& \quad 2.23 \,\, _{-\;\;\;\;0.32}^{+\;\;\;\;0.40}  \pm 0.22  \pm 0.13  \;  \% \, ,  \nonumber \\
&\mathcal{F}(\KsPiPi)^{\rm NR}                      &=& \quad 24.3 \pm  \,\,\,  1.3  \pm  \,\,\,  3.7   \pm \,\,\, 4.5 \;  \%  \, ,  \nonumber 
&\end{aligned}\]

\noindent 
where the statistical, experimental systematic and model uncertainties are split accordingly in that order.  
The results are in agreement with the measurements 
obtained by the \babar and \belle collaborations with decay-time-dependent flavour-tagged analyses~\cite{Dalseno:2008wwa,Aubert:2009me}, insofar as the DP model components can be compared.  
\par
The measurements of the \CP asymmetries are 

\[\begin{aligned}
&\mathcal{A}_{\CP}(K^{*}(892)^{-}\pi^{+})         &=& - 0.308 \pm  0.060  \pm 0.011  \pm 0.012    \, ,    \nonumber \\
&\mathcal{A}_{\CP}((K\pi)^{-}_0\pi^{+})            &=& - 0.032 \pm  0.047  \pm 0.016  \pm 0.027    \, ,    \nonumber \\
&\mathcal{A}_{\CP}(K^{*}_2(1430)^{-} \pi^{+}) &=& - 0.29  \phantom{2}  \pm    \,\,\,  0.22   \pm  \,\,\, 0.09    \pm  \,\,\, 0.03       \, ,   \nonumber \\
&\mathcal{A}_{\CP}(K^{*}(1680)^{-}\pi^{+})       &=& - 0.07  \phantom{2} \pm     \,\,\, 0.13    \pm  \,\,\, 0.02    \pm  \,\,\,0.03       \, ,   \nonumber \\
&\mathcal{A}_{\CP}(f_0(980)\KS)                        &=& \;\;\;  \;\, 0.28 \phantom{2} \pm  \,\,\, 0.27    \pm  \,\,\,0.05    \pm  \,\,\, 0.14       \, ,   \nonumber \\
&\end{aligned}\]
   
\noindent
where the uncertainties are statistical, experimental systematic and from the model.  
The statistical significance of having observed a nonvanishing \CP asymmetry in the decay $\overline{B}^0 \to K^*(892)^- \pi^+$, built from the likelihood ratio for the null hypothesis, is 6.7 
standard deviations and reduces to about 6 standard deviations taking into account the systematic uncertainties.  This measurement constitutes the first observation of a  \CP-violating 
asymmetry in the decay  $\overline{B}^0 \to K^*(892)^- \pi^+$.  The measured value is in good agreement with the world average $\mathcal{A}_{\CP}(K^{*}(892)^- \pi^+)   = -0.23 \pm 0.06$ 
with a similar precision.

\FloatBarrier

%
%
\section*{Acknowledgements}


\noindent We express our gratitude to our colleagues in the CERN
accelerator departments for the excellent performance of the LHC. We
thank the technical and administrative staff at the LHCb
institutes. We acknowledge support from CERN and from the national
agencies: CAPES, CNPq, FAPERJ and FINEP (Brazil); MOST and NSFC
(China); CNRS/IN2P3 (France); BMBF, DFG and MPG (Germany); INFN
(Italy); NWO (The Netherlands); MNiSW and NCN (Poland); MEN/IFA
(Romania); MinES and FASO (Russia); MinECo (Spain); SNSF and SER
(Switzerland); NASU (Ukraine); STFC (United Kingdom); NSF (USA).  We
acknowledge the computing resources that are provided by CERN, IN2P3
(France), KIT and DESY (Germany), INFN (Italy), SURF (The
Netherlands), PIC (Spain), GridPP (United Kingdom), RRCKI and Yandex
LLC (Russia), CSCS (Switzerland), IFIN-HH (Romania), CBPF (Brazil),
PL-GRID (Poland) and OSC (USA). We are indebted to the communities
behind the multiple open-source software packages on which we depend.
Individual groups or members have received support from AvH Foundation
(Germany), EPLANET, Marie Sk\l{}odowska-Curie Actions and ERC
(European Union), ANR, Labex P2IO, ENIGMASS and OCEVU, and R\'{e}gion
Auvergne-Rh\^{o}ne-Alpes (France), RFBR and Yandex LLC (Russia), GVA,
XuntaGal and GENCAT (Spain), Herchel Smith Fund, the Royal Society,
the English-Speaking Union and the Leverhulme Trust (United Kingdom).




\addcontentsline{toc}{section}{References}
\setboolean{inbibliography}{true}
\bibliographystyle{LHCb}
\bibliography{main,LHCb-PAPER,LHCb-CONF,LHCb-DP,LHCb-TDR}

\ifx\mcitethebibliography\mciteundefinedmacro
\PackageError{LHCb.bst}{mciteplus.sty has not been loaded}
{This bibstyle requires the use of the mciteplus package.}\fi
\providecommand{\href}[2]{#2}
\begin{mcitethebibliography}{10}
\mciteSetBstSublistMode{n}
\mciteSetBstMaxWidthForm{subitem}{\alph{mcitesubitemcount})}
\mciteSetBstSublistLabelBeginEnd{\mcitemaxwidthsubitemform\space}
{\relax}{\relax}

\bibitem{Christenson:1964fg}
J.~H. Christenson, J.~W. Cronin, V.~L. Fitch, and R.~Turlay,
  \ifthenelse{\boolean{articletitles}}{\emph{{Evidence for the $2\pi$ decay of
  the $K_2^0$ meson}},
  }{}\href{http://dx.doi.org/10.1103/PhysRevLett.13.138}{Phys.\ Rev.\ Lett.\
  \textbf{13} (1964) 138}\relax
\mciteBstWouldAddEndPuncttrue
\mciteSetBstMidEndSepPunct{\mcitedefaultmidpunct}
{\mcitedefaultendpunct}{\mcitedefaultseppunct}\relax
\EndOfBibitem
\bibitem{Aubert:2001nu}
BaBar collaboration, B.~Aubert {\em et~al.},
  \ifthenelse{\boolean{articletitles}}{\emph{{Observation of CP violation in
  the $B^0$ meson system}},
  }{}\href{http://dx.doi.org/10.1103/PhysRevLett.87.091801}{Phys.\ Rev.\ Lett.\
   \textbf{87} (2001) 091801},
  \href{http://arxiv.org/abs/hep-ex/0107013}{{\normalfont\ttfamily
  arXiv:hep-ex/0107013}}\relax
\mciteBstWouldAddEndPuncttrue
\mciteSetBstMidEndSepPunct{\mcitedefaultmidpunct}
{\mcitedefaultendpunct}{\mcitedefaultseppunct}\relax
\EndOfBibitem
\bibitem{Abe:2001xe}
Belle collaboration, K.~Abe {\em et~al.},
  \ifthenelse{\boolean{articletitles}}{\emph{{Observation of large CP violation
  in the neutral $B$ meson system}},
  }{}\href{http://dx.doi.org/10.1103/PhysRevLett.87.091802}{Phys.\ Rev.\ Lett.\
   \textbf{87} (2001) 091802},
  \href{http://arxiv.org/abs/hep-ex/0107061}{{\normalfont\ttfamily
  arXiv:hep-ex/0107061}}\relax
\mciteBstWouldAddEndPuncttrue
\mciteSetBstMidEndSepPunct{\mcitedefaultmidpunct}
{\mcitedefaultendpunct}{\mcitedefaultseppunct}\relax
\EndOfBibitem
\bibitem{HFAG}
Heavy Flavor Averaging Group, Y.~Amhis {\em et~al.},
  \ifthenelse{\boolean{articletitles}}{\emph{{Averages of $b$-hadron,
  $c$-hadron, and $\tau$-lepton properties as of summer 2016}},
  }{}\href{http://arxiv.org/abs/1612.07233}{{\normalfont\ttfamily
  arXiv:1612.07233}}, {updated results and plots available at
  \href{http://www.slac.stanford.edu/xorg/hflav/}{{\texttt{http://www.slac.stanford.edu/xorg/hflav/}}}}\relax
\mciteBstWouldAddEndPuncttrue
\mciteSetBstMidEndSepPunct{\mcitedefaultmidpunct}
{\mcitedefaultendpunct}{\mcitedefaultseppunct}\relax
\EndOfBibitem
\bibitem{Beneke:2003zv}
M.~Beneke and M.~Neubert, \ifthenelse{\boolean{articletitles}}{\emph{{QCD
  factorization for $B \to PP$ and $B \to PV$ decays}},
  }{}\href{http://dx.doi.org/10.1016/j.nuclphysb.2003.09.026}{Nucl.\ Phys.\
  \textbf{B675} (2003) 333},
  \href{http://arxiv.org/abs/hep-ph/0308039}{{\normalfont\ttfamily
  arXiv:hep-ph/0308039}}\relax
\mciteBstWouldAddEndPuncttrue
\mciteSetBstMidEndSepPunct{\mcitedefaultmidpunct}
{\mcitedefaultendpunct}{\mcitedefaultseppunct}\relax
\EndOfBibitem
\bibitem{Fleischer:2007mq}
R.~Fleischer, S.~Recksiegel, and F.~Schwab,
  \ifthenelse{\boolean{articletitles}}{\emph{{On puzzles and non-puzzles in $B
  \to \pi \pi$, $\pi K$ decays}},
  }{}\href{http://dx.doi.org/10.1140/epjc/s10052-007-0277-8}{Eur.\ Phys.\ J.\
  \textbf{C51} (2007) 55},
  \href{http://arxiv.org/abs/hep-ph/0702275}{{\normalfont\ttfamily
  arXiv:hep-ph/0702275}}\relax
\mciteBstWouldAddEndPuncttrue
\mciteSetBstMidEndSepPunct{\mcitedefaultmidpunct}
{\mcitedefaultendpunct}{\mcitedefaultseppunct}\relax
\EndOfBibitem
\bibitem{Baek:2009pa}
S.~Baek, C.-W. Chiang, and D.~London,
  \ifthenelse{\boolean{articletitles}}{\emph{{The $B \to \pi K$ puzzle: 2009
  update}}, }{}\href{http://dx.doi.org/10.1016/j.physletb.2009.03.062}{Phys.\
  Lett.\  \textbf{B675} (2009) 59},
  \href{http://arxiv.org/abs/0903.3086}{{\normalfont\ttfamily
  arXiv:0903.3086}}\relax
\mciteBstWouldAddEndPuncttrue
\mciteSetBstMidEndSepPunct{\mcitedefaultmidpunct}
{\mcitedefaultendpunct}{\mcitedefaultseppunct}\relax
\EndOfBibitem
\bibitem{Li:2009wba}
H.-n. Li and S.~Mishima, \ifthenelse{\boolean{articletitles}}{\emph{{Possible
  resolution of the $B \to \pi \pi, \pi K$ puzzles}},
  }{}\href{http://dx.doi.org/10.1103/PhysRevD.83.034023}{Phys.\ Rev.\
  \textbf{D83} (2011) 034023},
  \href{http://arxiv.org/abs/0901.1272}{{\normalfont\ttfamily
  arXiv:0901.1272}}\relax
\mciteBstWouldAddEndPuncttrue
\mciteSetBstMidEndSepPunct{\mcitedefaultmidpunct}
{\mcitedefaultendpunct}{\mcitedefaultseppunct}\relax
\EndOfBibitem
\bibitem{Khalil:2009zf}
S.~Khalil, A.~Masiero, and H.~Murayama,
  \ifthenelse{\boolean{articletitles}}{\emph{{$B \to K \pi$ puzzle and new
  sources of \CP violation in supersymmetry}},
  }{}\href{http://dx.doi.org/10.1016/j.physletb.2009.10.034}{Phys.\ Lett.\
  \textbf{B682} (2009) 74},
  \href{http://arxiv.org/abs/0908.3216}{{\normalfont\ttfamily
  arXiv:0908.3216}}\relax
\mciteBstWouldAddEndPuncttrue
\mciteSetBstMidEndSepPunct{\mcitedefaultmidpunct}
{\mcitedefaultendpunct}{\mcitedefaultseppunct}\relax
\EndOfBibitem
\bibitem{Cabibbo:1963yz}
N.~Cabibbo, \ifthenelse{\boolean{articletitles}}{\emph{Unitary symmetry and
  leptonic decays},
  }{}\href{http://dx.doi.org/10.1103/PhysRevLett.10.531}{Phys.\ Rev.\ Lett.\
  \textbf{10} (1963) 531}\relax
\mciteBstWouldAddEndPuncttrue
\mciteSetBstMidEndSepPunct{\mcitedefaultmidpunct}
{\mcitedefaultendpunct}{\mcitedefaultseppunct}\relax
\EndOfBibitem
\bibitem{Kobayashi:1973fv}
M.~Kobayashi and T.~Maskawa, \ifthenelse{\boolean{articletitles}}{\emph{{\CP}
  violation in the renormalizable theory of weak interaction},
  }{}\href{http://dx.doi.org/10.1143/PTP.49.652}{Prog.\ Theor.\ Phys.\
  \textbf{49} (1973) 652}\relax
\mciteBstWouldAddEndPuncttrue
\mciteSetBstMidEndSepPunct{\mcitedefaultmidpunct}
{\mcitedefaultendpunct}{\mcitedefaultseppunct}\relax
\EndOfBibitem
\bibitem{Buchalla:2005us}
G.~Buchalla, G.~Hiller, Y.~Nir, and G.~Raz,
  \ifthenelse{\boolean{articletitles}}{\emph{{The pattern of \CP asymmetries in
  \btos transitions}},
  }{}\href{http://dx.doi.org/10.1088/1126-6708/2005/09/074}{JHEP \textbf{09}
  (2005) 074}, \href{http://arxiv.org/abs/hep-ph/0503151}{{\normalfont\ttfamily
  arXiv:hep-ph/0503151}}\relax
\mciteBstWouldAddEndPuncttrue
\mciteSetBstMidEndSepPunct{\mcitedefaultmidpunct}
{\mcitedefaultendpunct}{\mcitedefaultseppunct}\relax
\EndOfBibitem
\bibitem{Grossman:1996ke}
Y.~Grossman and M.~P. Worah, \ifthenelse{\boolean{articletitles}}{\emph{{\CP
  asymmetries in \B decays with new physics in decay amplitudes}},
  }{}\href{http://dx.doi.org/10.1016/S0370-2693(97)00068-3}{Phys.\ Lett.\
  \textbf{B395} (1997) 241},
  \href{http://arxiv.org/abs/hep-ph/9612269}{{\normalfont\ttfamily
  arXiv:hep-ph/9612269}}\relax
\mciteBstWouldAddEndPuncttrue
\mciteSetBstMidEndSepPunct{\mcitedefaultmidpunct}
{\mcitedefaultendpunct}{\mcitedefaultseppunct}\relax
\EndOfBibitem
\bibitem{London:1997zk}
D.~London and A.~Soni, \ifthenelse{\boolean{articletitles}}{\emph{{Measuring
  the \CP angle $\beta$ in hadronic \btos penguin decays}},
  }{}\href{http://dx.doi.org/10.1016/S0370-2693(97)00695-3}{Phys.\ Lett.\
  \textbf{B407} (1997) 61},
  \href{http://arxiv.org/abs/hep-ph/9704277}{{\normalfont\ttfamily
  arXiv:hep-ph/9704277}}\relax
\mciteBstWouldAddEndPuncttrue
\mciteSetBstMidEndSepPunct{\mcitedefaultmidpunct}
{\mcitedefaultendpunct}{\mcitedefaultseppunct}\relax
\EndOfBibitem
\bibitem{Ciuchini:1997zp}
M.~Ciuchini {\em et~al.}, \ifthenelse{\boolean{articletitles}}{\emph{{\CP
  violating \B decays in the standard model and supersymmetry}},
  }{}\href{http://dx.doi.org/10.1103/PhysRevLett.79.978}{Phys.\ Rev.\ Lett.\
  \textbf{79} (1997) 978},
  \href{http://arxiv.org/abs/hep-ph/9704274}{{\normalfont\ttfamily
  arXiv:hep-ph/9704274}}\relax
\mciteBstWouldAddEndPuncttrue
\mciteSetBstMidEndSepPunct{\mcitedefaultmidpunct}
{\mcitedefaultendpunct}{\mcitedefaultseppunct}\relax
\EndOfBibitem
\bibitem{Dalitz:1953cp}
R.~H. Dalitz, \ifthenelse{\boolean{articletitles}}{\emph{{On the analysis of
  tau-meson data and the nature of the tau-meson}},
  }{}\href{http://dx.doi.org/10.1080/14786441008520365}{Phil.\ Mag.\
  \textbf{44} (1953) 1068}\relax
\mciteBstWouldAddEndPuncttrue
\mciteSetBstMidEndSepPunct{\mcitedefaultmidpunct}
{\mcitedefaultendpunct}{\mcitedefaultseppunct}\relax
\EndOfBibitem
\bibitem{Dalseno:2008wwa}
Belle collaboration, J.~Dalseno {\em et~al.},
  \ifthenelse{\boolean{articletitles}}{\emph{{Time-dependent Dalitz Plot
  measurement of \CP parameters in \BdtoKsPiPi decays}},
  }{}\href{http://dx.doi.org/10.1103/PhysRevD.79.072004}{Phys.\ Rev.\
  \textbf{D79} (2009) 072004},
  \href{http://arxiv.org/abs/0811.3665}{{\normalfont\ttfamily
  arXiv:0811.3665}}\relax
\mciteBstWouldAddEndPuncttrue
\mciteSetBstMidEndSepPunct{\mcitedefaultmidpunct}
{\mcitedefaultendpunct}{\mcitedefaultseppunct}\relax
\EndOfBibitem
\bibitem{Aubert:2009me}
\babar collaboration, B.~Aubert {\em et~al.},
  \ifthenelse{\boolean{articletitles}}{\emph{{Time-dependent amplitude analysis
  of \BdtoKsPiPi}},
  }{}\href{http://dx.doi.org/10.1103/PhysRevD.80.112001}{Phys.\ Rev.\
  \textbf{D80} (2009) 112001},
  \href{http://arxiv.org/abs/0905.3615}{{\normalfont\ttfamily
  arXiv:0905.3615}}\relax
\mciteBstWouldAddEndPuncttrue
\mciteSetBstMidEndSepPunct{\mcitedefaultmidpunct}
{\mcitedefaultendpunct}{\mcitedefaultseppunct}\relax
\EndOfBibitem
\bibitem{Nakahama:2010nj}
Belle collaboration, Y.~Nakahama {\em et~al.},
  \ifthenelse{\boolean{articletitles}}{\emph{{Measurement of \CP violating
  asymmetries in \BdtoKsKK decays with a time-dependent Dalitz approach}},
  }{}\href{http://dx.doi.org/10.1103/PhysRevD.82.073011}{Phys.\ Rev.\
  \textbf{D82} (2010) 073011},
  \href{http://arxiv.org/abs/1007.3848}{{\normalfont\ttfamily
  arXiv:1007.3848}}\relax
\mciteBstWouldAddEndPuncttrue
\mciteSetBstMidEndSepPunct{\mcitedefaultmidpunct}
{\mcitedefaultendpunct}{\mcitedefaultseppunct}\relax
\EndOfBibitem
\bibitem{Lees:2012kxa}
\babar collaboration, J.~P. Lees {\em et~al.},
  \ifthenelse{\boolean{articletitles}}{\emph{{Study of \CP violation in
  Dalitz-plot analyses of \decay{\Bd}{\Kp \Km \KS}, \decay{\Bp}{\Kp \Km \Kp},
  and \decay{\Bp}{\KS \KS \Kp}}},
  }{}\href{http://dx.doi.org/10.1103/PhysRevD.85.112010}{Phys.\ Rev.\
  \textbf{D85} (2012) 112010},
  \href{http://arxiv.org/abs/1201.5897}{{\normalfont\ttfamily
  arXiv:1201.5897}}\relax
\mciteBstWouldAddEndPuncttrue
\mciteSetBstMidEndSepPunct{\mcitedefaultmidpunct}
{\mcitedefaultendpunct}{\mcitedefaultseppunct}\relax
\EndOfBibitem
\bibitem{Aaij:2017zpx}
LHCb collaboration, R.~Aaij {\em et~al.},
  \ifthenelse{\boolean{articletitles}}{\emph{{Updated branching fraction
  measurements of $B^0_{(s)} \to K_{\mathrm{\scriptscriptstyle S}}^0 h^+
  h^{\prime -}$ decays}},
  }{}\href{http://dx.doi.org/10.1007/JHEP11(2017)027}{JHEP \textbf{11} (2017)
  027}, \href{http://arxiv.org/abs/1707.01665}{{\normalfont\ttfamily
  arXiv:1707.01665}}\relax
\mciteBstWouldAddEndPuncttrue
\mciteSetBstMidEndSepPunct{\mcitedefaultmidpunct}
{\mcitedefaultendpunct}{\mcitedefaultseppunct}\relax
\EndOfBibitem
\bibitem{Alves:2008zz}
LHCb collaboration, A.~A. Alves~Jr.\ {\em et~al.},
  \ifthenelse{\boolean{articletitles}}{\emph{{The \lhcb detector at the LHC}},
  }{}\href{http://dx.doi.org/10.1088/1748-0221/3/08/S08005}{JINST \textbf{3}
  (2008) S08005}\relax
\mciteBstWouldAddEndPuncttrue
\mciteSetBstMidEndSepPunct{\mcitedefaultmidpunct}
{\mcitedefaultendpunct}{\mcitedefaultseppunct}\relax
\EndOfBibitem
\bibitem{LHCb-DP-2014-002}
LHCb collaboration, R.~Aaij {\em et~al.},
  \ifthenelse{\boolean{articletitles}}{\emph{{LHCb detector performance}},
  }{}\href{http://dx.doi.org/10.1142/S0217751X15300227}{Int.\ J.\ Mod.\ Phys.\
  \textbf{A30} (2015) 1530022},
  \href{http://arxiv.org/abs/1412.6352}{{\normalfont\ttfamily
  arXiv:1412.6352}}\relax
\mciteBstWouldAddEndPuncttrue
\mciteSetBstMidEndSepPunct{\mcitedefaultmidpunct}
{\mcitedefaultendpunct}{\mcitedefaultseppunct}\relax
\EndOfBibitem
\bibitem{BBDT}
V.~V. Gligorov and M.~Williams,
  \ifthenelse{\boolean{articletitles}}{\emph{{Efficient, reliable and fast
  high-level triggering using a bonsai boosted decision tree}},
  }{}\href{http://dx.doi.org/10.1088/1748-0221/8/02/P02013}{JINST \textbf{8}
  (2013) P02013}, \href{http://arxiv.org/abs/1210.6861}{{\normalfont\ttfamily
  arXiv:1210.6861}}\relax
\mciteBstWouldAddEndPuncttrue
\mciteSetBstMidEndSepPunct{\mcitedefaultmidpunct}
{\mcitedefaultendpunct}{\mcitedefaultseppunct}\relax
\EndOfBibitem
\bibitem{Zemach:1965zz}
C.~Zemach, \ifthenelse{\boolean{articletitles}}{\emph{{Determination of the
  spins and parities of resonances}},
  }{}\href{http://dx.doi.org/10.1103/PhysRev.140.B109}{Phys.\ Rev.\
  \textbf{140} (1965) B109}\relax
\mciteBstWouldAddEndPuncttrue
\mciteSetBstMidEndSepPunct{\mcitedefaultmidpunct}
{\mcitedefaultendpunct}{\mcitedefaultseppunct}\relax
\EndOfBibitem
\bibitem{Williams:2010vh}
M.~Williams, \ifthenelse{\boolean{articletitles}}{\emph{{How good are your
  fits? Unbinned multivariate goodness-of-fit tests in high energy physics}},
  }{}\href{http://dx.doi.org/10.1088/1748-0221/5/09/P09004}{JINST \textbf{5}
  (2010) P09004}, \href{http://arxiv.org/abs/1006.3019}{{\normalfont\ttfamily
  arXiv:1006.3019}}\relax
\mciteBstWouldAddEndPuncttrue
\mciteSetBstMidEndSepPunct{\mcitedefaultmidpunct}
{\mcitedefaultendpunct}{\mcitedefaultseppunct}\relax
\EndOfBibitem
\bibitem{PDG2016}
Particle Data Group, C.~Patrignani {\em et~al.},
  \ifthenelse{\boolean{articletitles}}{\emph{{\href{http://pdg.lbl.gov/}{Review
  of particle physics}}},
  }{}\href{http://dx.doi.org/10.1088/1674-1137/40/10/100001}{Chin.\ Phys.\
  \textbf{C40} (2016) 100001}\relax
\mciteBstWouldAddEndPuncttrue
\mciteSetBstMidEndSepPunct{\mcitedefaultmidpunct}
{\mcitedefaultendpunct}{\mcitedefaultseppunct}\relax
\EndOfBibitem
\bibitem{ElBennich:2009da}
B.~El-Bennich {\em et~al.}, \ifthenelse{\boolean{articletitles}}{\emph{{$CP$
  violation and kaon-pion interactions in $B \to K \pi^+ \pi^-$ decays}},
  }{}\href{http://dx.doi.org/10.1103/PhysRevD.79.094005}{Phys.\ Rev.\
  \textbf{D79} (2009) 094005}, Erratum
  \href{http://dx.doi.org/10.1103/PhysRevD.83.039903}{ibid.\   \textbf{D83}
  (2011) 039903}, \href{http://arxiv.org/abs/0902.3645}{{\normalfont\ttfamily
  arXiv:0902.3645}}\relax
\mciteBstWouldAddEndPuncttrue
\mciteSetBstMidEndSepPunct{\mcitedefaultmidpunct}
{\mcitedefaultendpunct}{\mcitedefaultseppunct}\relax
\EndOfBibitem
\bibitem{Flatte:1976xu}
S.~M. Flatt\'e, \ifthenelse{\boolean{articletitles}}{\emph{{Coupled--channel
  analysis of the $\pi \eta$ and $K {\bar K}$ systems near $K \bar K$
  threshold}}, }{}\href{http://dx.doi.org/10.1016/0370-2693(76)90654-7}{Phys.\
  Lett.\  \textbf{B63} (1976) 224}\relax
\mciteBstWouldAddEndPuncttrue
\mciteSetBstMidEndSepPunct{\mcitedefaultmidpunct}
{\mcitedefaultendpunct}{\mcitedefaultseppunct}\relax
\EndOfBibitem
\bibitem{Muramatsu:2002jp}
CLEO collaboration, H.~Muramatsu {\em et~al.},
  \ifthenelse{\boolean{articletitles}}{\emph{{Dalitz analysis of $D^0 \to \KS
  \pi^+ \pi^-$}},
  }{}\href{http://dx.doi.org/10.1103/PhysRevLett.89.251802}{Phys.\ Rev.\ Lett.\
   \textbf{89} (2002) 251802},
  \href{http://arxiv.org/abs/hep-ex/0207067}{{\normalfont\ttfamily
  arXiv:hep-ex/0207067}}\relax
\mciteBstWouldAddEndPuncttrue
\mciteSetBstMidEndSepPunct{\mcitedefaultmidpunct}
{\mcitedefaultendpunct}{\mcitedefaultseppunct}\relax
\EndOfBibitem
\bibitem{Gounaris:1968mw}
G.~J. Gounaris and J.~J. Sakurai,
  \ifthenelse{\boolean{articletitles}}{\emph{{Finite-width corrections to the
  vector-meson dominance prediction for $\rho \to e^+ e^-$}},
  }{}\href{http://dx.doi.org/10.1103/PhysRevLett.21.244}{Phys.\ Rev.\ Lett.\
  \textbf{21} (1968) 244}\relax
\mciteBstWouldAddEndPuncttrue
\mciteSetBstMidEndSepPunct{\mcitedefaultmidpunct}
{\mcitedefaultendpunct}{\mcitedefaultseppunct}\relax
\EndOfBibitem
\bibitem{Ablikim:2004wn}
BES collaboration, M.~Ablikim {\em et~al.},
  \ifthenelse{\boolean{articletitles}}{\emph{{Resonances in $J / \psi \to \phi
  \pi^+ \pi^-$ and $\phi K^+ K^-$}},
  }{}\href{http://dx.doi.org/10.1016/j.physletb.2004.12.041}{Phys.\ Lett.\
  \textbf{B607} (2005) 243},
  \href{http://arxiv.org/abs/hep-ex/0411001}{{\normalfont\ttfamily
  arXiv:hep-ex/0411001}}\relax
\mciteBstWouldAddEndPuncttrue
\mciteSetBstMidEndSepPunct{\mcitedefaultmidpunct}
{\mcitedefaultendpunct}{\mcitedefaultseppunct}\relax
\EndOfBibitem
\bibitem{LHCb-PAPER-2017-10}
LHCb Collaboration, R.~Aaij {\em et~al.},
  \ifthenelse{\boolean{articletitles}}{\emph{{Updated branching fraction
  measurements of $B^0_{(s)} \to K_{\mathrm{\scriptscriptstyle S}}^0 h^+
  h^{\prime -}$ decays}},
  }{}\href{http://dx.doi.org/10.1007/JHEP11(2017)027}{JHEP \textbf{1711} (2017)
  027. 42 p}\relax
\mciteBstWouldAddEndPuncttrue
\mciteSetBstMidEndSepPunct{\mcitedefaultmidpunct}
{\mcitedefaultendpunct}{\mcitedefaultseppunct}\relax
\EndOfBibitem
\bibitem{LHCb-PAPER-2013-018}
LHCb collaboration, R.~Aaij {\em et~al.},
  \ifthenelse{\boolean{articletitles}}{\emph{{First observation of $\CP$
  violation in the decays of $\Bs$ mesons}},
  }{}\href{http://dx.doi.org/10.1103/PhysRevLett.110.221601}{Phys.\ Rev.\
  Lett.\  \textbf{110} (2013) 221601},
  \href{http://arxiv.org/abs/1304.6173}{{\normalfont\ttfamily
  arXiv:1304.6173}}\relax
\mciteBstWouldAddEndPuncttrue
\mciteSetBstMidEndSepPunct{\mcitedefaultmidpunct}
{\mcitedefaultendpunct}{\mcitedefaultseppunct}\relax
\EndOfBibitem
\bibitem{LHCb-PAPER-2012-009}
LHCb collaboration, R.~Aaij {\em et~al.},
  \ifthenelse{\boolean{articletitles}}{\emph{{Measurement of the $\Dsp$--$\Dsm$
  production asymmetry in $7$\tev $\proton\proton$ collisions}},
  }{}\href{http://dx.doi.org/10.1016/j.physletb.2012.06.001}{Phys.\ Lett.\
  \textbf{B713} (2012) 186},
  \href{http://arxiv.org/abs/1205.0897}{{\normalfont\ttfamily
  arXiv:1205.0897}}\relax
\mciteBstWouldAddEndPuncttrue
\mciteSetBstMidEndSepPunct{\mcitedefaultmidpunct}
{\mcitedefaultendpunct}{\mcitedefaultseppunct}\relax
\EndOfBibitem
\bibitem{Ko:2010mk}
B.~R. Ko, E.~Won, B.~Golob, and P.~Pakhlov,
  \ifthenelse{\boolean{articletitles}}{\emph{{Effect of nuclear interactions of
  neutral kaons on \CP asymmetry measurements}},
  }{}\href{http://dx.doi.org/10.1103/PhysRevD.84.111501}{Phys.\ Rev.\
  \textbf{D84} (2011) 111501},
  \href{http://arxiv.org/abs/1006.1938}{{\normalfont\ttfamily
  arXiv:1006.1938}}\relax
\mciteBstWouldAddEndPuncttrue
\mciteSetBstMidEndSepPunct{\mcitedefaultmidpunct}
{\mcitedefaultendpunct}{\mcitedefaultseppunct}\relax
\EndOfBibitem
\bibitem{LHCb-PUB-2011-025}
M.~De~Cian {\em et~al.},
  \ifthenelse{\boolean{articletitles}}{\emph{{Measurement of the track finding
  efficiency}}, }{}
  \href{http://cdsweb.cern.ch/search?p=LHCb-PUB-2011-025&f=reportnumber&action_search=Search&c=LHCb+Notes}
  {LHCb-PUB-2011-025}\relax
\mciteBstWouldAddEndPuncttrue
\mciteSetBstMidEndSepPunct{\mcitedefaultmidpunct}
{\mcitedefaultendpunct}{\mcitedefaultseppunct}\relax
\EndOfBibitem
\bibitem{LHCb-DP-2012-004}
R.~Aaij {\em et~al.}, \ifthenelse{\boolean{articletitles}}{\emph{{The \lhcb
  trigger and its performance in 2011}},
  }{}\href{http://dx.doi.org/10.1088/1748-0221/8/04/P04022}{JINST \textbf{8}
  (2013) P04022}, \href{http://arxiv.org/abs/1211.3055}{{\normalfont\ttfamily
  arXiv:1211.3055}}\relax
\mciteBstWouldAddEndPuncttrue
\mciteSetBstMidEndSepPunct{\mcitedefaultmidpunct}
{\mcitedefaultendpunct}{\mcitedefaultseppunct}\relax
\EndOfBibitem
\bibitem{Aston:1987ir}
D.~Aston {\em et~al.}, \ifthenelse{\boolean{articletitles}}{\emph{{A study of
  $K^- \pi^+$ scattering in the reaction $K^- p \to K^- \pi^+ n$ at 11~GeV/c}},
  }{}\href{http://dx.doi.org/10.1016/0550-3213(88)90028-4}{Nucl.\ Phys.\
  \textbf{B296} (1988) 493}\relax
\mciteBstWouldAddEndPuncttrue
\mciteSetBstMidEndSepPunct{\mcitedefaultmidpunct}
{\mcitedefaultendpunct}{\mcitedefaultseppunct}\relax
\EndOfBibitem
\end{mcitethebibliography}

\newpage


 
\newpage
\centerline{\large\bf LHCb collaboration}
\begin{flushleft}
\small
R.~Aaij$^{40}$,
B.~Adeva$^{39}$,
M.~Adinolfi$^{48}$,
Z.~Ajaltouni$^{5}$,
S.~Akar$^{59}$,
J.~Albrecht$^{10}$,
F.~Alessio$^{40}$,
M.~Alexander$^{53}$,
A.~Alfonso~Albero$^{38}$,
S.~Ali$^{43}$
% ...

{\footnotesize \it
% ...
$ ^{5}$Clermont Universit{\'e}, Universit{\'e} Blaise Pascal, CNRS/IN2P3, LPC, Clermont-Ferrand, France\\
% ...
$ ^{10}$Fakult{\"a}t Physik, Technische Universit{\"a}t Dortmund, Dortmund, Germany\\
% ...
$ ^{38}$ICCUB, Universitat de Barcelona, Barcelona, Spain\\
$ ^{39}$Universidad de Santiago de Compostela, Santiago de Compostela, Spain\\
$ ^{40}$European Organization for Nuclear Research (CERN), Geneva, Switzerland\\
% ...
$ ^{43}$Nikhef National Institute for Subatomic Physics, Amsterdam, The Netherlands\\
% ...
$ ^{48}$H.H. Wills Physics Laboratory, University of Bristol, Bristol, United Kingdom\\
% ...
$ ^{53}$School of Physics and Astronomy, University of Glasgow, Glasgow, United Kingdom\\
% ...
$ ^{59}$University of Cincinnati, Cincinnati, OH, United States\\
% ...
}
\end{flushleft}



\end{document}